\documentclass[a4paper]{elsarticle}

\usepackage[utf8x]{inputenc}
\usepackage[english]{babel}
\usepackage[T1]{fontenc}
\usepackage{graphicx}
\usepackage{listings}
\usepackage[dvipsnames]{xcolor}
\usepackage{paralist}
\usepackage{caption}
\usepackage{subcaption}
\usepackage{url}
\usepackage{inconsolata}
\usepackage[ruled,vlined]{algorithm2e} 
\usepackage{amsmath} 
\usepackage{a4wide} 

\lstdefinelanguage{somd} {
    captionpos=b,
    basicstyle={\ttfamily\scriptsize},
    numbers=left,
    tabsize=2,
    keywordstyle=\sffamily\bfseries,
    commentstyle=\color{ForestGreen},
    language=Java,
    frame=single,
    breaklines=true,
	showstringspaces=false,
	breaklines=true,                
	breakatwhitespace=true,        
	frame=tb,
	texcl,
	morekeywords = {atomic, dist, shared, barrier, reduce, self, sync, distshared}
}

\DeclareMathOperator{\norm}{norm}

\newcommand{\kwd}[1]{\textsf{\texttt{#1}}\xspace}

\newcommand{\distk}{\kwd{dist}}
\newcommand{\reducek}{\kwd{reduce}}
\newcommand{\sharedk}{\kwd{shared}}

\newcommand{\selfk}{\kwd{self}}
\newcommand{\synck}{\kwd{sync}}
\newcommand{\singlek}{\kwd{single}}

\journal{Special issue of Journal of Computer and System Sciences}

\begin{document}

 \begin{frontmatter}

\title{Heterogeneous Programming with Single Operation Multiple Data\tnoteref{label0}\tnoteref{label1}}

\tnotetext[label0]{Preprint of the article accepted for publication in a special issue of Elsevier's Journal of Computer and System Sciences dedicated to the 14th IEEE International Conference on High Performance Computing and Communication (HPCC 2012)}
\tnotetext[label1]{This work was partially funded by FCT-MEC in the framework of the  PEst-OE/EEI/UI0527/2011 - Centro de Inform\'atica e Tecnologias da Informa\c{c}\~ao  (CITI/FCT/UNL) - 2011-2012.}

\author{Herv\'e Paulino}
 \ead{herve.paulino@fct.unl.pt}
 \ead[url]{http://asc.di.fct.unl.pt/~herve}

\author{Eduardo Marques}
\ead{eduardo$\_$raf@gmail.com}
 
\address{CITI / Departamento de Inform\'atica \\
Faculdade de Ci\^encias e Tecnologia, Universidade Nova de Lisboa\\
2829-516 Caparica, Portugal}

\begin{abstract}
Heterogeneity is omnipresent in today's commodity computational systems, which 
comprise
at least one multi-core Central Processing Unit (CPU) and  one Graphics Processing Unit (GPU).
Nonetheless, all this computing power is not being harnessed
in mainstream computing, as the programming of these systems
entails many details of the underlying architecture and of its distinct execution models.
Current research  on parallel programming is addressing these issues
but, still, the system's heterogeneity is exposed at language level.  

This paper proposes a uniform framework, grounded on  the Single Operation Multiple Data model,  for the programming of such heterogeneous systems.
The model is declarative,  empowering the compiler to generate code for multiple architectures from the same source. 
To this extent, we designed  a simple extension of the Java programming language that embodies the model, and developed
a compiler that generates code for both multi-core CPUs and GPUs.
%
A performance evaluation attests the validity of the approach that, despite being based on a simple programming model,  is able to deliver    performance gains on par with hand-tuned data parallel multi-threaded Java applications.

\end{abstract}


\end{frontmatter}

\section{Introduction}

The landscape of  computing systems has altered in the last few years, with 
the shift from frequency to core scaling in  CPU design, and the increasing popularity of General Purpose computation on GPUs (GPGPU) \cite{gpgpu}.
The architecture  of current commodity computational systems is quite   complex and heterogeneous,   featuring a combination of, at least, one multi-core CPU and one GPU. 
This is further aggravated by the distinct nature of 
the architectural and execution models in place.

Therefore, the programming  of these heterogeneous systems as a whole raises several challenges:
which computations to run on  each kind of processing unit; 
how to decompose a problem to fit the execution model of the target processing unit;
how to map this decomposition in the system's complex memory hierarchy;
among others.
Tackling them efficiently requires true knowledge of parallel computing and computer architecture.
However, mainstream software developers do not wish to deal with
 details, inherent to the underlying architecture, that are completely abstracted in classic sequential and even concurrent programming.
 Consequently, most of the available computing power is not really exploited. 
 
 The definition of high-level programming models for heterogeneous computing
 has been the driver of a considerable amount of recent research.
Existing languages, such as    X10 \cite{x10-gpus}, Chapel \cite{chapel-gpu}, and
  StreamIt \cite{streamit-gpu} have incorporated GPU support, and new languages, such as Lime \cite{lime}, have been proposed altogether. 
 Of these works, we are particularly interested in X10 and  Chapel, since,  to the best of our knowledge, they are the only to target 
 heterogeneous systems at node and cluster level.
 However, as will be detailed in Section \ref{sec:related}, their approach is not platform independent, exposing details of the target architecture at language level.

In this paper, we propose the use of the Single Operation Multiple Data  (SOMD) model, presented in \cite{somd-hpcc},
to provide a uniform framework for the programming of  this range of architectures.
SOMD introduces the expression of data parallelism at subroutine 
level.
The calling  of a subroutine in this context spawns several tasks, 
each operating on a separate partition of the input dataset.
These  tasks are offloaded for parallel execution
by multiple workers,
and run  in conformity to a variation of the Single Program Multiple Data (SPMD) execution model.

This approach provides a framework  for the average programmer to express data parallel computations by annotating unaltered sequential subroutines,
hence
taking advantage of the parallel nature of the   target hardware without  having to program specialized code.
In \cite{somd-hpcc} we  debated that this approach is viable for the programming of both shared and distributed memory architectures.
In this extended version, we broad this claim to the GPU computing field, thus addressing the issue of heterogeneous computing.
To this extent, our contributions are: 
\begin{inparaenum}[i)]
\item a more detailed presentation of the SOMD  execution model and a refinement  of its programming model, featuring new constructs; 
\item the conceptual realization of this execution model on shared memory architectures,  distributed memory architectures, and  GPU accelerated systems;
\item the effective compilation process for tackling heterogeneous computing, namely targeting multi-core CPUs and GPUs;
\item the evaluation of the code generated by our current prototype against hand-tuned data parallel multi-threaded applications.
\end{inparaenum}

The remainder of this paper is structured as follows:
the next section provides a detailed motivation for expressing  data parallelism at subroutine level, when compared to the usual loop-level  parallelism.
We also make evidence that approaches such as X10 and Chapel are bound to the underlying architecture.
Section  \ref{sec:somd} overviews the SOMD execution and programming models.
Section \ref{sec:multiple-archs} presents our conceptual realization of the model on multiple architectures.
Sections \ref{sec:comp} and \ref{sec:java}  describe, respectively, the compilation process for 
generating code for both shared memory multi-core CPUs and GPUs, and the required runtime support.
Section \ref{sec:eval} evaluates our prototype implementation from a performance perspective.
Finally, Section \ref{sec:conclusions} presents our concluding remarks.

\section{Background and Related Work}
\label{sec:related}

Data-parallelism is traditionally expressed at loop-level in both shared and distributed memory programming.
Regarding the former, OpenMP \cite{openmp} is the most popular parallel computing framework.
It provides a mix of compiler directives, library calls and environment variables, being that 
data-parallelism is expressed by annotating with directives the loops suitable for parallel execution.
More recently, similar approaches have made their way into GPGPU.  The most notorious example is OpenACC \cite{openacc}, a directive-based specification for offloading computation to GPUs and managing the associated data transfers.

Another representative  system for shared-memory parallelism is Cilk \cite{cilk}, a C  language extension that offers  primitives to spawn and synchronize concurrent tasks (C functions).
In Cilk,  data-parallelism is expressed by programming a loop to spawn the desired number of tasks, each receiving as argument the boundaries of the subset of the problem's domain it will work upon. A C++ derivation of the extension was recently  integrated in Intel's parallel programming toolkit \cite{cilkpp}. This derivation adds some new features to the system, among which a
special loop construct (\texttt{cilk\_for}) for loop-level parallelism.

Intel Threading Building Blocks (TBB) \cite{tbb} is a C++  template library that factorizes  recurring parallel patterns.
Data-parallelism in TBB may be expressed through the specialization of  loop templates, supplying the task's body and, eventually, a partitioning strategy.

Loop-level data parallelism also prevails in distributed memory environments.
dipSystem \cite{dips} applied Cilk-like parallelism to distributed environments,  providing a uniform interface for the programming of both shared and distributed memory architectures. For that purpose, the spawned tasks could be parametrized with the data to work upon and operate over
explicitly managed shared variables.
Single Program Multiple Data (SPMD) languages, such as UPC \cite{upc},
 require a special \texttt{forall} construct to express loops 
 that work upon data distributed across multiple nodes.
 In the particular case of UPC (and Co-Array Fortran \cite{caf}) the Partitioned Global Addressing Space (PGAS) model provides the means for the programmer to explore the affinity between data and computation and hence reduce the communication overhead.

X10 \cite{x10} extends the PGAS model with notion of asynchronous activity, providing a framework for both task and data parallelism.
Data distribution is performed at runtime upon a set of places (abstractions of network nodes).
In this context, data parallelism is expressed through loops that iterate over this set of places, working only on the data residing locally at each place.
Inter-place parallelism is expressed much in the same way as in the original Cilk.
More recently, the  language has also been extended to support the offload of computation to GPUs  \cite{x10-gpus}, which are presented as sub-places of the original place, the node hosting the GPU.
This approach adequately exposes the isolation of the GPU's memory relatively to node's main memory.
However, the X10 programming model is very imperative,  forcing the programmer to be also aware of the underlying execution model.
In order to be suitable for GPU execution an X10 asynchronous task must begin with two 
\texttt{for} loops: one to denote the distribution of the work per thread-groups, and a second to denote 
the  distribution of the work within a group.
Moreover, the enclosed code must fulfil  several restrictions, such as the absence of method invocations.

\begin{lstlisting}[float, caption={Vector addition - X10  version for shared memory},label=lst:x10-mc, language=somd, morekeywords={def,val,finish, var, in, async, atomic,here}]
def vectorAdd(a:Array[Int](1), b:Array[Int](1), numThreads:Int): Array[Int](1) {
 var c = new Array[Int](a.size);
 val mySize = a.size/numThreads;
 finish { // Synchronization construct
  for (p in 0..(numThreads-1)) 
   async { // Spawn asynchronous activity
    for (i in (p*mySize)..((p+1)*mySize)) c(i) = a(i) + b(i);  }
 }
 return c;
}
\end{lstlisting}

\begin{lstlisting}[float, caption={Vector addition - X10  version for cluster environments},label=lst:x10-dm, language=somd, language=java, morekeywords={at,def,val,finish, var, in, async, atomic,here}]
def vectorAddCluster(a:DistArray[Int](1), b:DistArray[Int](1), numThreads:Int): DistArray[Int](1)  {
 var c = DistArray.make(a.dist); // Distribute array
 finish { // Synchronization construct
  for (p in a.dist.places()) // For each place
    async at(p) { // Spawn asynchronous activity at p
     vectorAddSM(c|here, a|here, b|here, numThreads);  } // Local computation
 }
 return c;
}  
\end{lstlisting}

\begin{lstlisting}[float, caption={Vector addition - X10  version for GPUs},label=lst:x10-gpu, language=somd, morekeywords={def,val,finish, var, in, async, atomic,here,at,@CUDA}]
def vectorAddGPU(a:Array[Int](1), b:Array[Int](1), Place gpu): Array[Int](1)  {
 at (gpu) @CUDA {
  val blocks = CUDAUtilities.autoBlocks(); // Atomically define the number of blocks
  val threads = CUDAUtilities.autoThreads(); // Atomically define the number of threads
  var c = new Array[Float](a.length);  // Result array
  finish for ([b] in 0..blocks-1) async { // Distribute work across the blocks
   var c_shm = new Array[Int](a.length); // Local data, shared by all threads in a block
   finish for ([t] in 0..threads-1) async { // Distribute work across the threads
     c_shm(t) = a(t) + b(t); // Implicit data transfer of a and b
   }
   finish Array.asyncCopy(c, c_shm, a.length); // Copy data from GPU to host
 }
 return c;
} 
\end{lstlisting}

Since X10 embraces some of the same goals of this work, we illustrate how it can be utilized to implement the simple problem of adding two vectors in 
 shared memory, distributed memory, and  GPUs (Listings \ref{lst:x10-mc} to \ref{lst:x10-gpu})\footnote{For the sake of simplicity we assume that both vectors have the same size.}.
Note that the code  entails specificities of the underlying architectures, reducing abstraction and disabling portability.
Furthermore, there is no support for partitioning the contents of an existing data-structure.
One must first allocate a distributed data-structure and then copy the contents from the local to the distributed version.
For simplicity's sake, in the cluster version we assume that this task has been previously taken care of, and thus the method receives two distributed arrays.
 In Listing \ref{lst:x10-dm}, the $|$ operator, the \texttt{here}
constant  and the \texttt{at} construct denote, respectively, domain restriction,  current place, and computation locality.
Method \texttt{vectorAddSM} in that same listing implements the behaviour of \texttt{vectorAdd} (Listing \ref{lst:x10-mc}), but with the target array passed as argument.

\begin{lstlisting}[float, caption={Vector addition - Chapel  version for shared memory},label=lst:chapel-mc, language=somd, morekeywords={proc,in,dmapped, const, var, forall}]
proc vectorAddSM(A:[] int, B:[] int): [] int {
 var C:  [a.domain] int; 
 forall (a,b,c) in (A,B,C) do  c = a + b;
 return C;     
}
\end{lstlisting}
	
\begin{lstlisting}[float, caption={Vector addition - Chapel  version for cluster environments},label=lst:chapel-dm ,language=somd, morekeywords={real,proc,dmapped, const, var, forall}]
proc vectorAddCluster(A:[] int, B:[] int): [] int {
 const space = [1..A.size] dmapped Block(boundingBox=[1..A.size]); 
 var DA, DB, DC : [space] int; // Distribute vectors
 DA = A; // Load A into DA 
 DB = B; // Load B into DB 
 forall (a,b,c) in (DA,DB,DC) do 
   c = a + b;  // Cluster wide computation, performed locally at each location of (a,b,c)
 return C;    
}
\end{lstlisting}

\begin{lstlisting}[float, caption={{Vector addition - Chapel  version for GPUs}},label=lst:chapel-gpu, language=somd, morekeywords={in,proc,dmapped, const, var, forall}]
proc vectorAddGPU(A:[] int, B:[] int): [] int {
 const space = [1..A.size] dmapped GPUDist(rank=1); 
 var DA, DB, DC : [space] int; // Allocate space for the vectors i	n the GPU
 DA = A; // Load A into DA 
 DB = B; // Load B into DB 
 forall (a,b,c) in (DA,DB,DC) do
   c = a + b; // Implicit copy of DA and DB
 return DC;   // Return result vector, implicit copy from the GPU's memory
}
\end{lstlisting}

Chapel \cite{chapel} is also a PGAS language
that embodies the  shared memory, distributed memory and GPU programming paradigms \cite{chapel-gpu}. 
It borrows many concepts of ZPL \cite{zpl}, a fact that has a clear impact on its high-level constructs for data parallelism.
A \texttt{forall} loop iterates over a domain's index set or over an array, spawning one thread per available processing unit in multi-core architectures.
On GPUs, each loop iteration corresponds to a GPU thread.
There are still asymmetries between the three versions presented in Listings \ref{lst:chapel-mc} to \ref{lst:chapel-gpu}, namely in the expressing of how the data must be partitioned among the existing addressing spaces in cluster and GPU environments.
Moreover, as for X10, there is no support for distributing the contents of an existing data-structure.

MapReduce \cite{mapreduce}   is a  programming model for  distributed data-parallel processing of large datasets.  
  Parallelism is not expressed at loop-level. Instead, the programmer must structure the input dataset in terms of key-value pairs
   and  the computation as a sequence of two stages:  \textit{map} and \textit{reduce}.
   The \textit{map} stage defines a transformation to be applied in parallel to the entire input data-set.
   Its output is  subsequently  aggregated and combined by the \textit{reduce} stage to produce the computation's final result.
  Additionally, most implementations provide  hooks for user-defined functions, which  can be  specified to  provide application specific strategies relating to the management of intermediate data. Otherwise, the runtime system performs all  execution steps automatically.  
MapReduce is essentially used in cluster environments, however there also implementations for shared memory \cite{phoenix} and GPUs \cite{mars,gpumr}.

\begin{lstlisting}[float, caption=Vector addition - OpenCL version, label=lst:opencl,language=somd, morekeywords={__kernel,__global, unsigned}]
__kernel void vectorAdd( __global int *a, __global int *b,  __global int *c, const unsigned int len)  {  
   int i = get_global_id(0); // Obtain the thread's global identifier
   if (i < len) // Make sure that the index is not outside the vector's boundaries
     c[i] = a[i] + b[i];                                 
}                                                             
\end{lstlisting}

Finally, the base GPGPU programming model exported by OpenCL \cite{opencl} and CUDA \cite{cuda}
differs  considerably from the loop-level parallelism.
The computation is divided into two categories: 
GPU devices run  computational \textit{kernels} that follow the SPMD execution model, while 
\textit{host} computations, processed by the CPU,  have the purpose of  orchestrating and issuing
 these device executions.
Programming in  such APIs is a challenging exercise, since not only do the kernels need to be structured according to the SPMD model, but also, the host has to oversee many low-level programming concerns. The latter range from memory/resource management, to performing data transfers between host and device memories, or even synchronizing with the device.
Since our work also targets GPU devices, we depict in 
Listing \ref{lst:opencl}  a kernel for the computation of the  vector addition example. The \texttt{{\_\_global}} keyword specifies that the data is located in the GPU's global memory, and the \texttt{get\_global\_id()} function retrieves the thread's global identifier in the defined one-dimensional grid.
The host counterpart of the example  is omitted due to its verbosity, approximately 100 lines.

These base GPGPU frameworks are  ideal for squeezing the full potential of a GPU, but will never make their way into more mainstream programming.
They can  almost be seen as a portable assembly for GPGPU.
Moreover, even though OpenCL code is portable between CPUs and GPUs, and there are even some proposals that tackle cluster environments \cite{dopencl},  performance portability across this range of architectures is all but  trivial. 
The degree of abstraction is much lower than the one offered by Chapel and X10.

On the other hand, the latter cannot deliver the same level of performance on GPUs.
A key issue is the efficient management of data movement between host and device.
With this limitation in mind, both Chapel and X10 enable the programmer to take explicit control upon all data transfer operations to, and from, the  GPU.
The amount of extra code  is not particularly significant, but the level of abstraction is greatly reduced.

\section{The SOMD Execution Model}
\label{sec:somd}

\begin{figure}
\centering
\includegraphics[width=12cm]{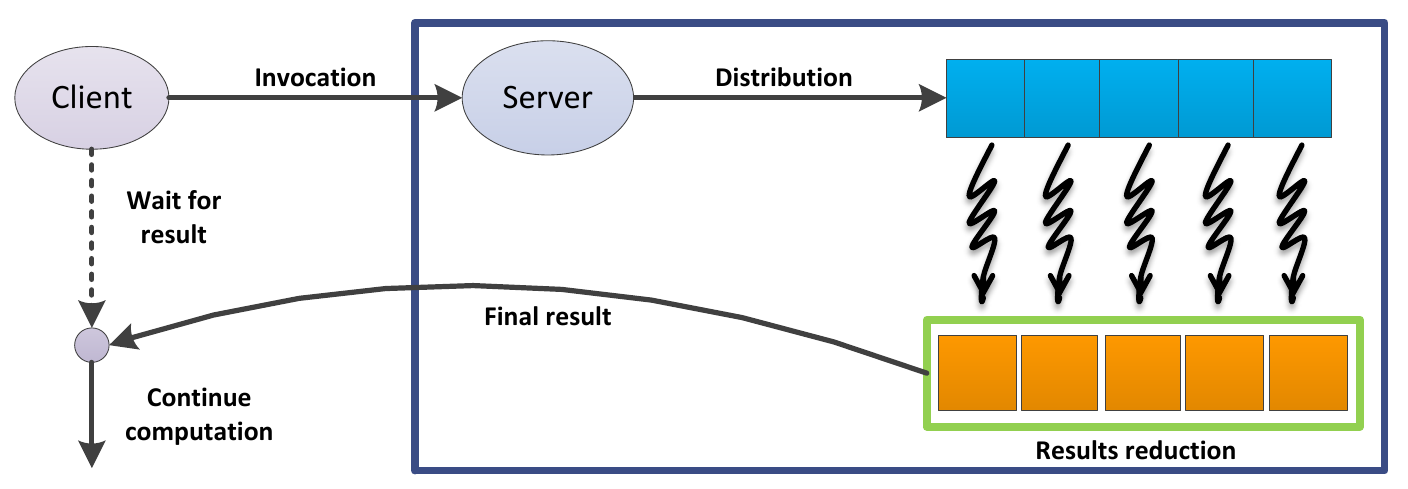}
\caption{Invocation and execution of a SOMD method}
\label{img:invoc}
\end{figure}

The SOMD execution model  consists on carrying out  multiple instances of a given method\footnote{Our current research is applying the SOMD model to object-oriented languages, thus, from this point onward, we will use the method terminology  rather than subroutine.} in 
parallel, over different partitions of the input dataset.
The invocation is decoupled from the execution, a characteristic  that makes the parallel nature of the execution model transparent to the invoker.
 As is illustrated in Figure \ref{img:invoc}, 
the invocation is synchronous, complying with the common semantics of subroutine invocation in most imperative programming languages, such a C and Java.
The execution stage is carried out  by multiple concurrent  flows, each operating over one of  the partitions of the input dataset.
If each of these  method instances (MIs) produces a result, their collection will be fed to a reduction stage that 
will use them to compute the method's final result.

The model is presented to the programmer in the form of a \textit{Distribute-Map-Reduce} (DMR) paradigm
that shares some ideas with the MapReduce paradigm.
A brief description of each stage follows:
\begin{description}

\item[{Distribute}] partitions the target value into a collection of values of the same type. 
It can be applied  to multiple  input arguments and local variables.

\item[{Map}]  applies a MI to each partition of the input dataset.

\item[{Reduce}] combines the partial results of the previous stage to compute the method's final result.

\end{description}

\begin{figure}
\centering
\includegraphics[width=12cm]{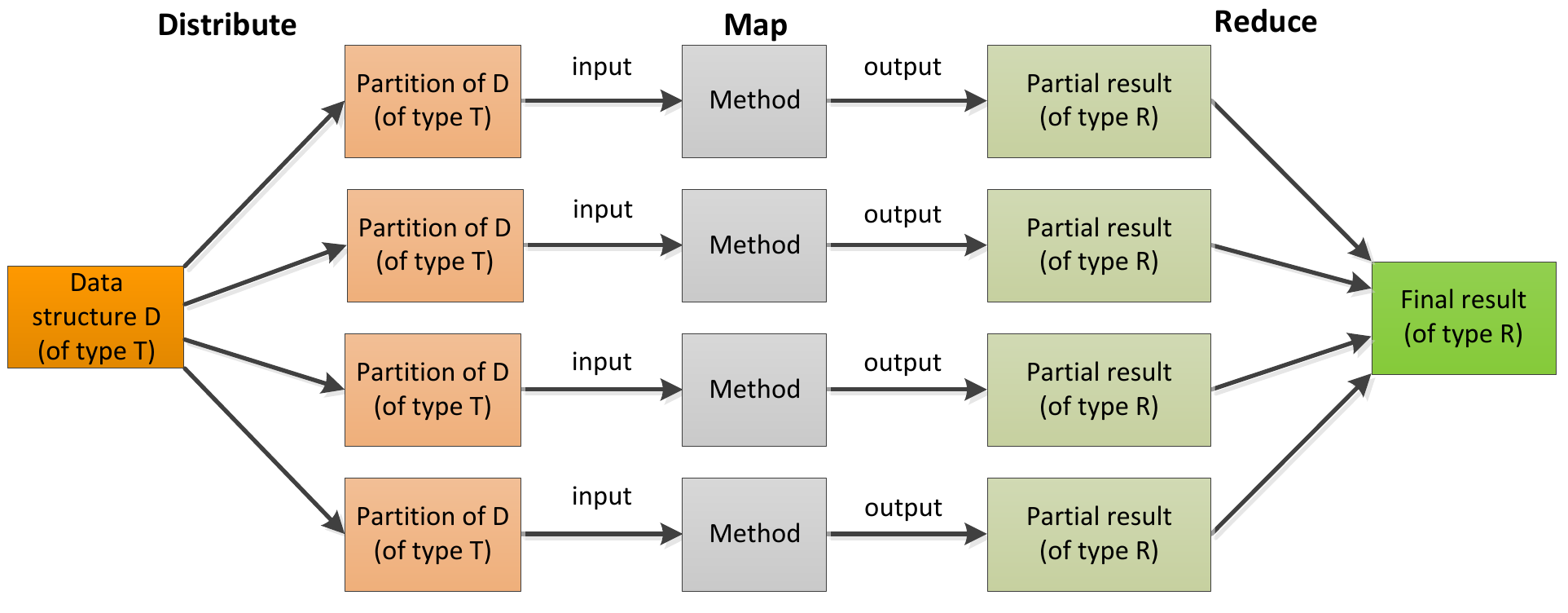}
\caption{The  \textit{Distribute-Map-Reduce} (DMR) paradigm applied to a method with prototype \texttt{R method(T D)}}
\label{img:model}
\end{figure}

Note that the prototype of the original method  does not have to be modified, in order to be suitable for the application 
of the	DMR paradigm.
Given an argument of type $T$, a distribution over such argument must be a 
 function of type \[T \mapsto List<T>\]
Moreover,  a reduction applied to a method that returns a value of type $R$
must be a function of type \[List<R>\	 \mapsto R\]
Thus, the method's application complies to its original prototype, since it receives
one of the elements of the distribution set (of type $T$) and provides an output  
of type \textit{R}. 
Figure \ref{img:model}  depicts a simplified version of the paradigm comprising a single distributed value. 

\subsection{Programming Constructs}
The  constructs presented in this subsection extend and refine the ones originally proposed in \cite{somd-hpcc}, namely with built-in, self and intermediate reductions.  
We have prototyped our proposal in the Java language, 
hence the exhibited programming examples   will comply with the Java syntax.
To provide a linguistic framework  as little intrusive as possible, we have extended the language with four simple constructs.
These extensions could be performed through the language's annotation framework.
However,  the latter does not possess the expressive power we require, particularly in the annotation of blocks of code.

\paragraph{Partitioning and Reduction Strategies}
The application of a partitioning strategy is expressed through the \distk qualifier, which can target  
both method parameters and local variables.
Unlike regular Java qualifiers, \distk may be parametrized,  for instance, with the name of the class implementing the partitioner and its arguments.
Partitioner implementations (that we will refer to as \textit{strategies}) must comply to a pre-determined interface and can be  user-defined.
However, since the large majority of data parallel computations work upon arrays, we will treat these data-structures  specially, providing built-in implementations.
By default, we assume a block-partitioning strategy.

Reduction operations are denoted by the \reducek qualifier and have a method-wide scope.
Similarly to partitions, these may be parametrized with the name of the implementing class and its arguments.
Moreover, once again, built-in strategies are supplied:
(a) reduction through primitive operations (currently $+, -$, and $*$) can be written simply as \reducek($op$), 
e.g. \reducek(+),
and (b) the 
assembling of  partially computed arrays is assumed by default whenever the method's return value is an array. In such cases,  the \reducek qualifier may be omitted. 
Currently, reductions are sequentially and deterministically applied  to the list of results output by the map stage.
Commutativity should not be an issue, but there are  situations that require reductions to be associative (see Section \ref{sec:archs:cluster}).
Our prototype implementation does not validate any of these properties. Thus it is up to the programmer to ensure them.

Listing \ref{lst:vectoradd} illustrates the SOMD implementation of  the addition of two vectors.
Counterposing with the X10 and Chapel versions depicted in Section \ref{sec:related}, the method's body is the same of the sequential implementation and 
the constructs required for the parallel execution are much more declarative.
The programmer is aware that multiple method instances  will  execute in parallel, possibly in different locations with their own addressing spaces, but no details of the underlying architecture are exposed at language level.
This approach gives way   for the compiler to generate code for distinct architectural designs from a single implementation.

\begin{lstlisting}[float, caption=Vector addition - SOMD version, language= somd, label=lst:vectoradd]
int[] vectorAdd(dist int[] a, dist int[] b) {
  int[] c  = new int[a.length];  
  for (int i = 0 ; i < a.length; i++)  
    c[i] = a[i] + b[i];
  return c;   
}
\end{lstlisting}

\begin{lstlisting}[float, caption=Sum of the elements of an array, label=lst:arraysum,language=somd]
reduce(self)
int  sum(dist int[] a) {
  int sum = 0;
  for (int i = 0 ; i < a.length; i++) 
    sum += a[i]; 
  return sum; 
}
\end{lstlisting}

\paragraph{Self-Reductions}
A recurrent pattern in SOMD programming is the application of the same behaviour on both the map
and the  reduction stages.
Given this, a second built-in distribution, \selfk, natively provides such behaviour.
Listing \ref{lst:arraysum} applies it  in the sum of the elements of a vector.
Both the map and the reduction stages will execute instances of the \texttt{sum} method.
Note that there are no visible data-races, since \texttt{sum} is local to each method instance and to the reduction itself.
If, eventually, the reduction is performed in parallel, it is up to the compiler to generate the concurrency management code.
Note that in this particular case  \reducek(+) could also be used.

\begin{figure}
\centering
\includegraphics[width=12cm]{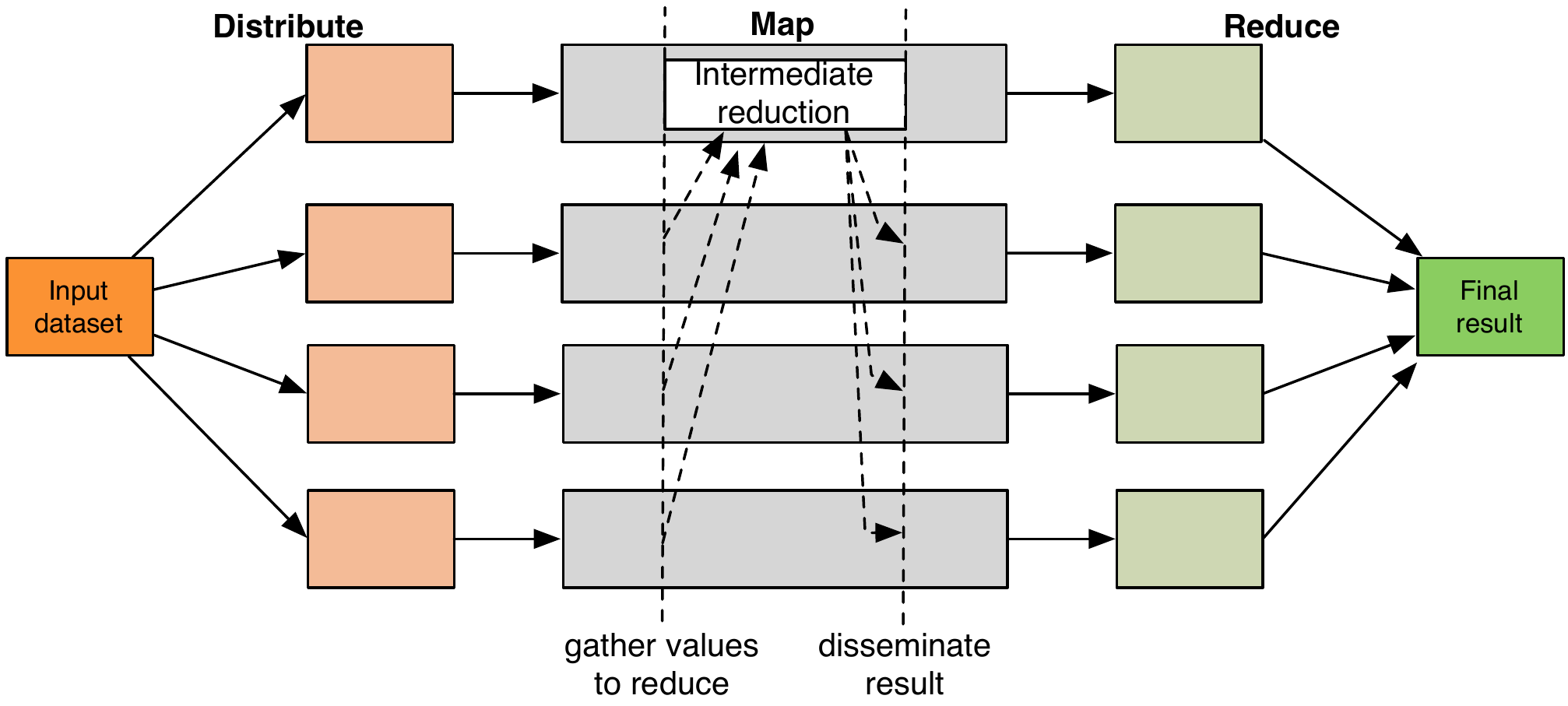}
\caption{Intermediate reductions in the DMR paradigm}
\label{fig:interred}
\end{figure}

\paragraph{Intermediate Reductions}
A SOMD method may invoke an auxiliary method.
If the latter method performs a reduction, such operation is applied to the results computed by all MIs. 
The motivation behind this feature is to, on one hand, 
allow the nesting of SOMD methods, and, on the other, enable multiple
reductions over a single distributed dataset. This prevents distributing the same data more than once, and therefore narrows the communication overhead.
Currently MI divergence in this context is not supported, meaning that 
nested SOMD invocations cannot be conditional.
Figure \ref{fig:interred} illustrates how intermediate reductions integrate with the DMR paradigm.
One of the MIs assumes the responsibility of  computing the operation. Ergo it must receive the operation's input values and 
disseminate the computed result to the remainder MIs.

Listing  \ref{lst:norm} illustrates the use of an intermediate reduction to compute the norm of a vector:
$\norm(a_1, a_2, ..., a_n) = \sqrt{a_1^2 + a_2^2 + \cdots + a_n^2}$.
Method \texttt{sumProd} computes the sum of the products of each partition, reducing the result through the + reduction.
The square-root is computed locally by each MI (line 2) to be later used in the normalization (line 3).
Lastly, the default reduction for arrays assembles the result vector.

\begin{lstlisting}[float, caption=Vector normalization, label=lst:norm,numbers=right, language=somd]
int[] norm(dist int[] a) {
 double norm  = Math.sqrt(sumProd(a));
 for (int i = 0 ; i < a.length; i++) a[i] = a[i]/norm; 
 return a; 
}

reduce(+) 
double sumProd(int[] a) {
 double sumProd = 0;
 for (int i = 0 ; i < a.length; i++) sumProd += a[i]*a[i];
 return sumProd;
}
\end{lstlisting}

\paragraph{Shared Array Positions}
From the programmer's perspective, communication between MIs 
can also be performed through shared memory. It is up to the compiler to generate the necessary remote communication (if necessary).
For that purpose, we allow for both array positions and scalars objects
to be shared.
We abide to the PGAS model in the sense that the programmer must explicitly state which data is remotely accessible, i.e., does not reside in a MIs local memory.

\begin{figure}
\centering
\begin{subfigure}{0.45\textwidth}
	\centering
	\includegraphics[width=4cm]{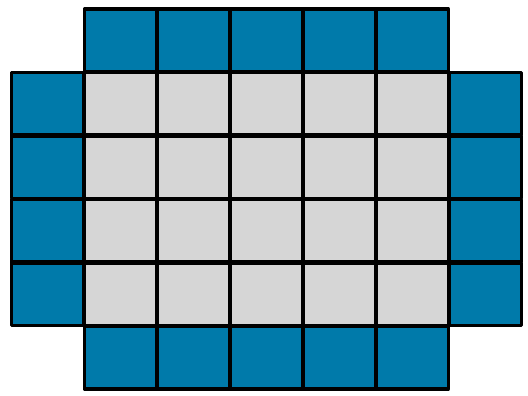} 
	\caption{\texttt{view} = $<1,1>,<1,1>$}
\end{subfigure}
\begin{subfigure}{0.45\textwidth}
	\centering
	\includegraphics[width=4cm]{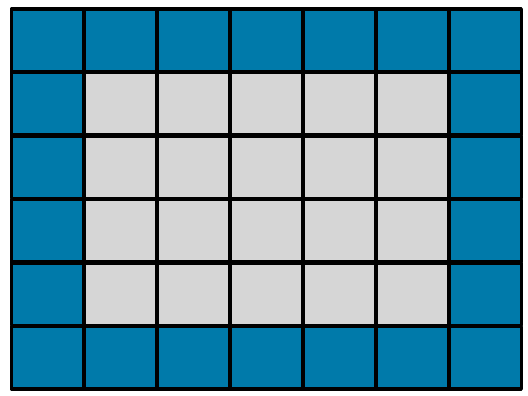} 
	\caption{\texttt{polyview} = $<1,1>,<1,1>$ }
\end{subfigure}
\caption{The gray mesh denotes the elements assigned to the MI, while the blue rows and columns denote the ones it is able to access from its neighbours.}
\label{fig:sap}
\end{figure}

\begin{lstlisting}[float, caption=A stencil computation inspired by  SOR, label=lst:sor,  language=somd]
reduce(+)
double stencil(dist(view = <1,1>, <1,1>) double[][] G, int num_iterations) {
 double Gtotal = 0;
 for (int p = 0; p < num_iterations; p++) 
  sync { 
   for (int i = 1; i < G.length-1; i++) 
    for (int j = 1; j < G[0].length-1; j++)             
      G[i][j] = (G[i-1][j]+G[i+1][j]+G[i][j-1]+G[i][j+1]) + a_constant * G[i][j];
  }
  
  for (int i = 1; i < G.length-1; i++) 
   for (int j = 1; j < G[0].length-1; j++)   
    Gtotal += G[i][j];
  return Gtotal;          
}
\end{lstlisting}

Regarding distributed arrays,
 we define an extra argument to the \distk construct: \texttt{view}, 
a vector that, for each dimension, indicates how many indexes beyond (but adjacent to) 
the  boundaries of the partition assigned to a given MI are visible to the latter. 
Consider a SOMD implementation of a stencil computation inspired in  the Successive Over-Relaxation (SOR) 
benchmark taken from the JavaGrande suite \cite{javagrande}, depicted in Listing \ref{lst:sor}.
The view vector \[<1,1>,<1,1>\] present in line 1 indicates that for dimensions 1 and 2 the MI is able to expand its view in 1 position in either direction, as illustrated in Figure \ref{fig:sap} (a). 
Polygonal views, such as the one in Figure \ref{fig:sap} (b), are expressed through a different keyword: \texttt{polyview}.
This view concept is somehow reminiscent of ZPL's region borders \cite{zpl}.
 Note that by default a matrix is partitioned into two-dimensional blocks.
 Parameter \texttt{dim} allows for the explicit specification on  which dimension(s) to partition.

Performance issues dictate that, in distributed memory architectures, memory consistency should be relaxed.
Although we do not compromise ourselves with any particular consistency model, we provide language support for such relaxation.
We introduce a memory-fence synchronization construct, \synck, that, when in place,
forces all MIs to have the same view of a particular variable or of all shared memory (if no particular variable is given)
once the enclosed code has completed its execution.
It can be interpreted as a data-centred version of X10's \texttt{finish}.
\texttt{finish} ensures that all asynchronous computations spawned within the scope of the construct complete
before the execution flow may proceed,  while \synck ensures 
that the memory is consistent across all MIs before the execution flow may proceed.
Iterative computing,  such as the one depicted in Lisiting \ref{lst:sor},  is a paradigmatic setting for the display of \synck's usefulness.
The iterations of the loop at line 6 are data-dependent - 
 the contents of the \texttt{G} matrix computed by 
an iteration are required by the following. 
Ergo, we enclose them in a \synck block.

\paragraph{Shared scalars}
We also allow scalars to be shared between MIs, through the \sharedk qualifier.
As with shared arrays, consistency is enforced through \synck blocks.
However, with shared variables this can only be achieved if the local copies are reduced into a single global value.
For that purpose, \synck blocks must be combined with reduction functions.
This, in fact, is nothing more than syntactic sugar for an intermediate reduction.
Listing \ref{lst:norm2} showcases the use of shared variables in a new version of the vector normalization problem.

In addition, we allow reductions to be applied directly upon a \synck block targeting a single distributed value.

\begin{lstlisting}[float, caption=Vector normalization (version 2) , label=lst:norm2,language=somd]
int[] normalize(dist int[] a) {
 shared double norm = 0;
 sync reduce(+) (norm)  {
   for (int i = 0 ; i < a.length; i++) norm += a[i]*a[i]; // local operation
 } // all copies of norm are combined to produce an identical copy in all MIs
 norm = Math.sqrt(norm);
 for (int i = 0 ; i < a.length; i++) a[i] = a[i]/norm; 
 return a; 
}
\end{lstlisting}

\section{Supporting Multiple Architectures}
\label{sec:multiple-archs}

Despite its simplicity, the SOMD execution model can  effectively   run computations on 
a wide range of architectures, multi-core CPUs, GPUs and  clusters of the former.
This flexibility arises from the declarative nature of the proposed linguistic constructs.
No architecture  specific details are encoded at language level, a characteristic that enables the portability of the source code, empowering the compiler and the runtime system.
The programmer  still detains control of where the computation must take place, but this information is segregated from the functional concerns.

In order for the code to be fully portable across such disparate set of architectures, we must apply a  restriction:
parameters must be used for input  only.
\textit{Inout} parameters are common in shared memory programming, but not suitable for distributed memory environments.
Therefore, any result produced by a SOMD method must be explicitly returned.

The SOMD model induces  a master-worker pattern.
The master is responsible for: 
\begin{inparaenum}[i)]
\item the application of the partitioning strategy over the original dataset;
\item the dispatching of the MIs to the slave workers\footnote{At this point, for the sake of simplicity, we are only executing  the \textit{map} stage in parallel (the execution of the MIs).
Nonetheless, parallelism may also be applied to both  the partitioning and reduction stages.};
\item the collection of the partial results, and;
\item the computation of   the reduction stage.
\end{inparaenum}
In turn,  a slave executes one or more MIs on a target architecture.
These roles may be mixed up, as the master may itself execute a subset of the MIs it distributes, assuming the role of a slave, and,  
as will be detailed in Subsection \ref{sec:archs:cluster}, the application of 
a hierarchical work distribution strategy may lead a slave to further decompose the received dataset, assuming a mixed  role of master.

The master-slave pattern is a common paradigm for parallel computing in all  our target architectures, and thus can be efficiently implemented.
However,  its concrete realization  may differ considerably from one architecture to another.

\subsection{Shared-memory Architectures}
\label{sec:archs:sm}

\begin{figure}
\centering
\includegraphics[width=\linewidth]{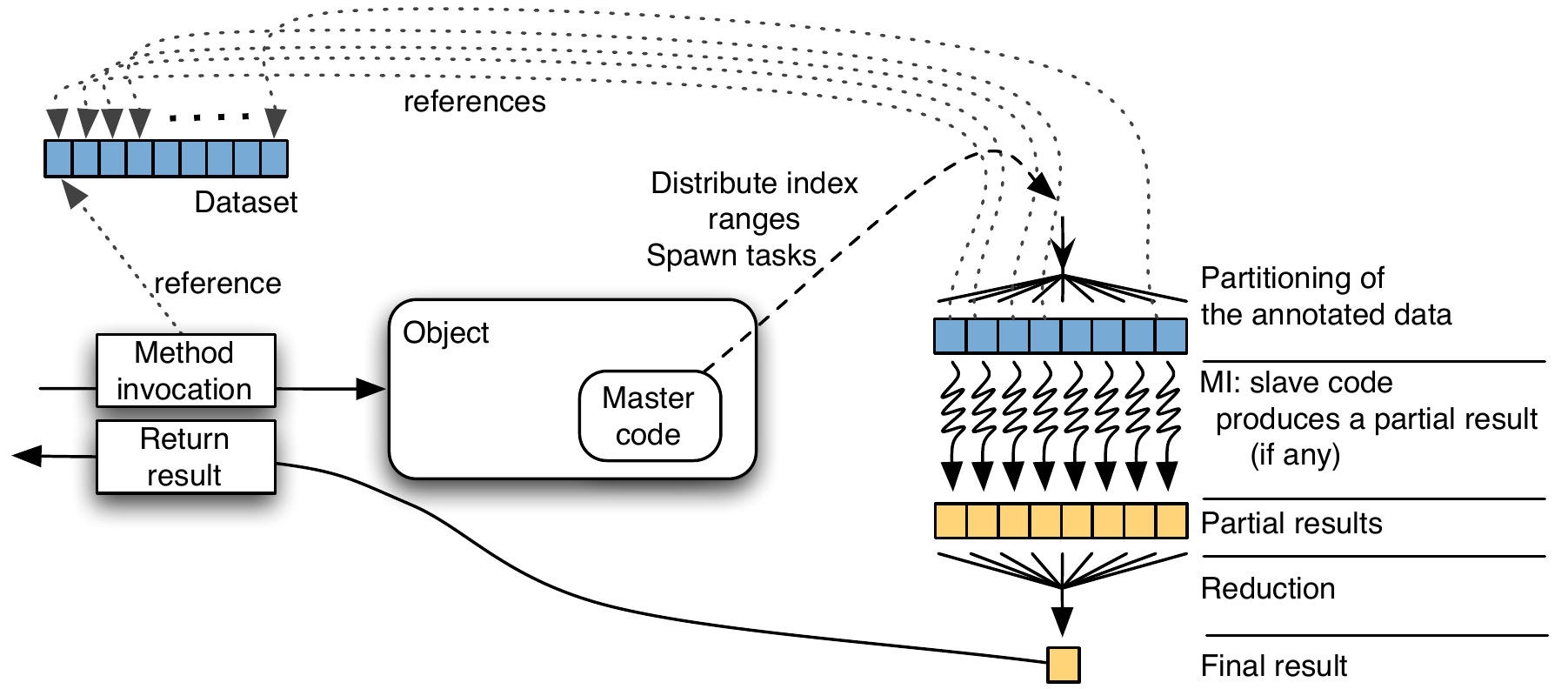}
\caption{SOMD execution model on shared memory}
\label{fig:sm}
\end{figure}

Shared memory architectures are the simplest ones to deal with.
As proposed in \cite{somd-hpcc}, the set of slaves can be realized through a pool of threads.
However, as concluded in that same paper, effectively splitting  data in Java-like languages can be too much of a burden 
for shared-memory programming, as the  splitting process requires the creation of new objects and the subsequent copy of  data.
Consequently, copy-free approaches are welcomed. 
For instance, a simple distribution of index ranges over  arrays is preferable to the actual partitioning of the array's contents. 
The number of MIs to spawn may follow different criteria, such as one per available processor, or 
as many as required to have the associated data partition fit a given level of the cache hierarchy.
Figure \ref{fig:sm} provides an overall perspective of the realization of the model in shared memory architectures.

In what concerns memory consistency across MIs.  In shared-memory, we assume a strict memory model, hence consistency is trivially obtained with a synchronization barrier at the end of the code enclosed by a \synck block (more details in  Subsection \ref{sec:comp:sm}).

\subsection{Clusters}
\label{sec:archs:cluster}

In cluster environments,  distributed arrays must be 
effectively scattered across the participating nodes.
This operation can be carried out hierarchically, since  distribution strategies are intrinsically associative.
Ergo, a straightforward approach is to simply  split the data, as evenly as possible, among the target nodes and 
then perform the same operation inside the node, by distributing index ranges among the existing slaves, as described in the previous section.
The handling of heterogeneity requires more sophisticated partitioning and load-balancing algorithms, such as hierarchical work-stealing \cite{Hws}.
These concerns, however crucial for the efficient implementation of SOMD-like distribution in heterogeneous  environments, are outside the scope of this paper.

Reductions can also be performed hierarchically, as a mean to decrease the amount of data transferred to the master.
However, the associativity assumptions deduced in the partitioning stage
are not generically valid for reduction operations.
Programmers are obliged to supply associative reduction operations, whose property 
may be statically verified at cluster deployment-time. 

\begin{figure}
\centering
\includegraphics[width=10cm]{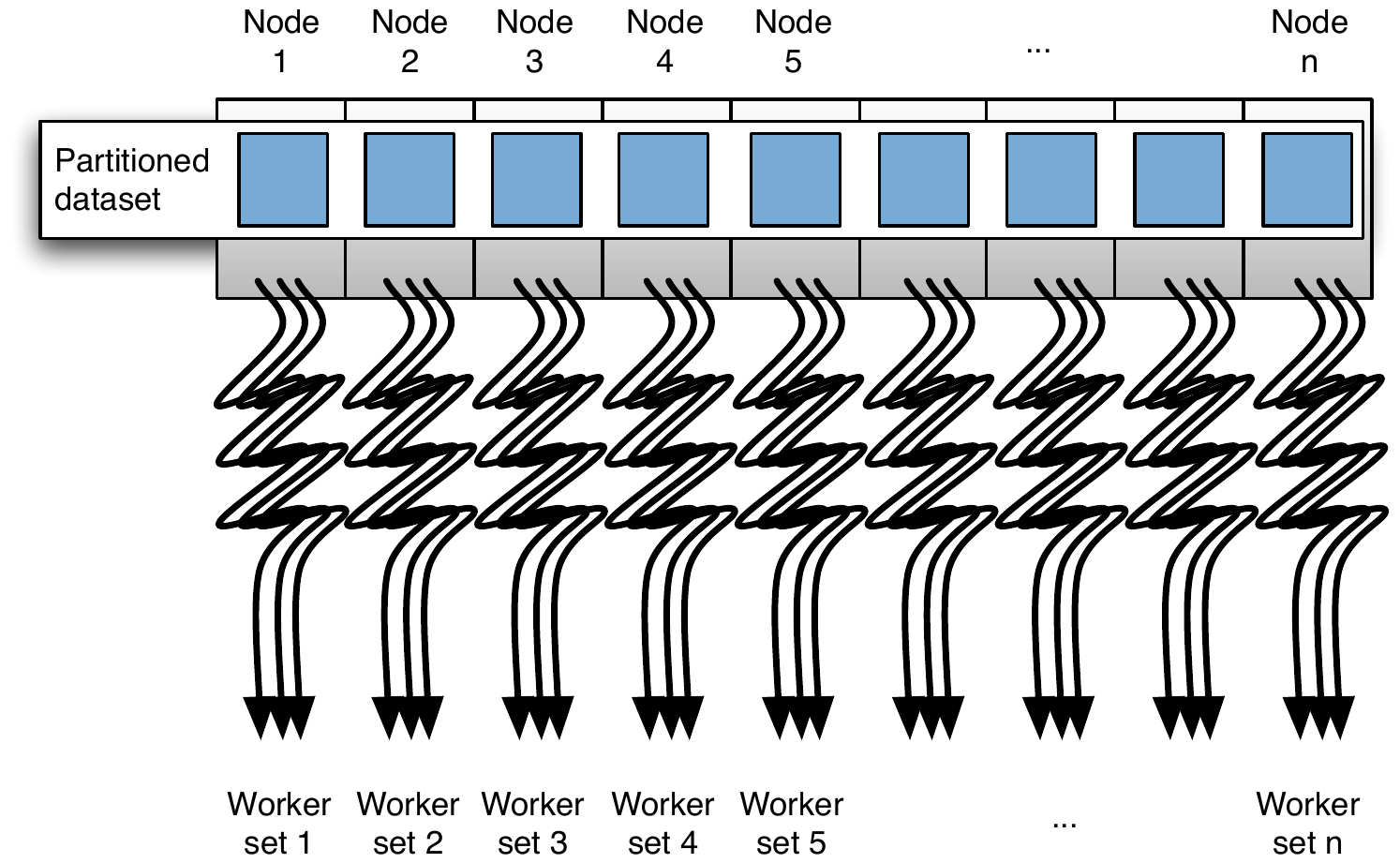}
\caption{PGAS properties of the SOMD model}
\label{fig:pgas}
\end{figure}

\begin{figure}
\centering
\includegraphics[width=\linewidth]{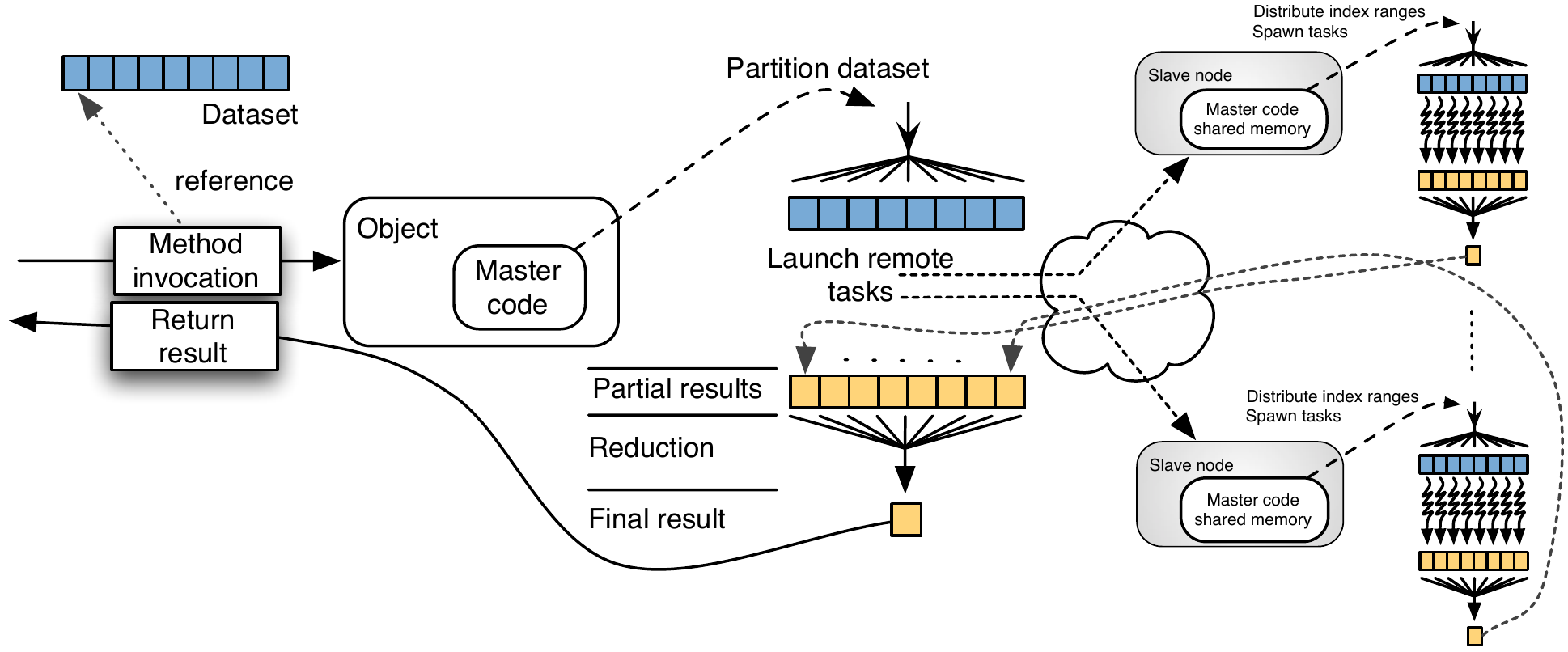}
\caption{SOMD execution model on distributed memory architectures}
\label{fig:dm}
\end{figure}

This overall   approach embraces the concepts of hierarchical data-parallelism popularized by Sequoia \cite{sequoia} and subsequently used in several other systems/languages.
An interesting property of the SOMD model is that it embodies PGAS properties by design (Figure \ref{fig:pgas}).
Unless explicitly expressed, each MI operates on local data.
This contrasts with existing  approaches where 
remote communication is assumed by default and it is affinity that must explicitly expressed, e.g.
the domain restriction operator in X10 (Listing \ref{lst:x10-dm}) and the fourth parameter of UPC's \texttt{forall}  loop. 
Figure \ref{fig:dm} illustrates  the hierarchy of the execution model in distributed memory environments.

When it comes to distributed shared arrays,  
each node may hold sub-parts of the array visible to remotely executing MIs.
The efficient implementation of such a system requires a runtime service with characteristics akin to 
one-way messaging services, such as GasNet \cite{gasnet}. To implement such support in the Java language is still a challenge to surmount.  
Finding out where the data is can be easily achieved by computing a hash code for	 the index.
Naturally, caching and weak consistency models are welcomed to reduce communication overhead.

In this paper, we will not address this class of architectures, however  
some initial results have been presented in \cite{elina}. This substantiates the viability of the execution model in cluster environments.

\subsection{Graphic Processing Units}
\label{sec:archs:gpu}

Like SOMD,  GPGPU  is grounded on the SPMD model. 
Data-parallel computations 
are carried out by a set of threads organized in blocks, 
each running the same code.
Therefore, it is not difficult to establish a parallel between the SOMD and the GPGPU models, 
an evidence that sustains our claim to use SOMD  for  GPU programming.

Both CUDA and OpenCL divide the computation in two categories: host and device. The host computations orchestrate and issue device executions, whilst 
the device runs the parallel computations: the kernel functions.
Once again,  similarities with the SOMD execution model are evident.
The host may execute the master side of a SOMD method execution, whilst the device executes the kernel(s) 
needed to realize the execution of the MIs.
The code generated for the master side is  similar to its generated counterpart   for cluster environments:
it must partition the input dataset, launch the MIs, perform the reduction stage, and return the result.

\begin{figure}
\includegraphics[width=\linewidth]{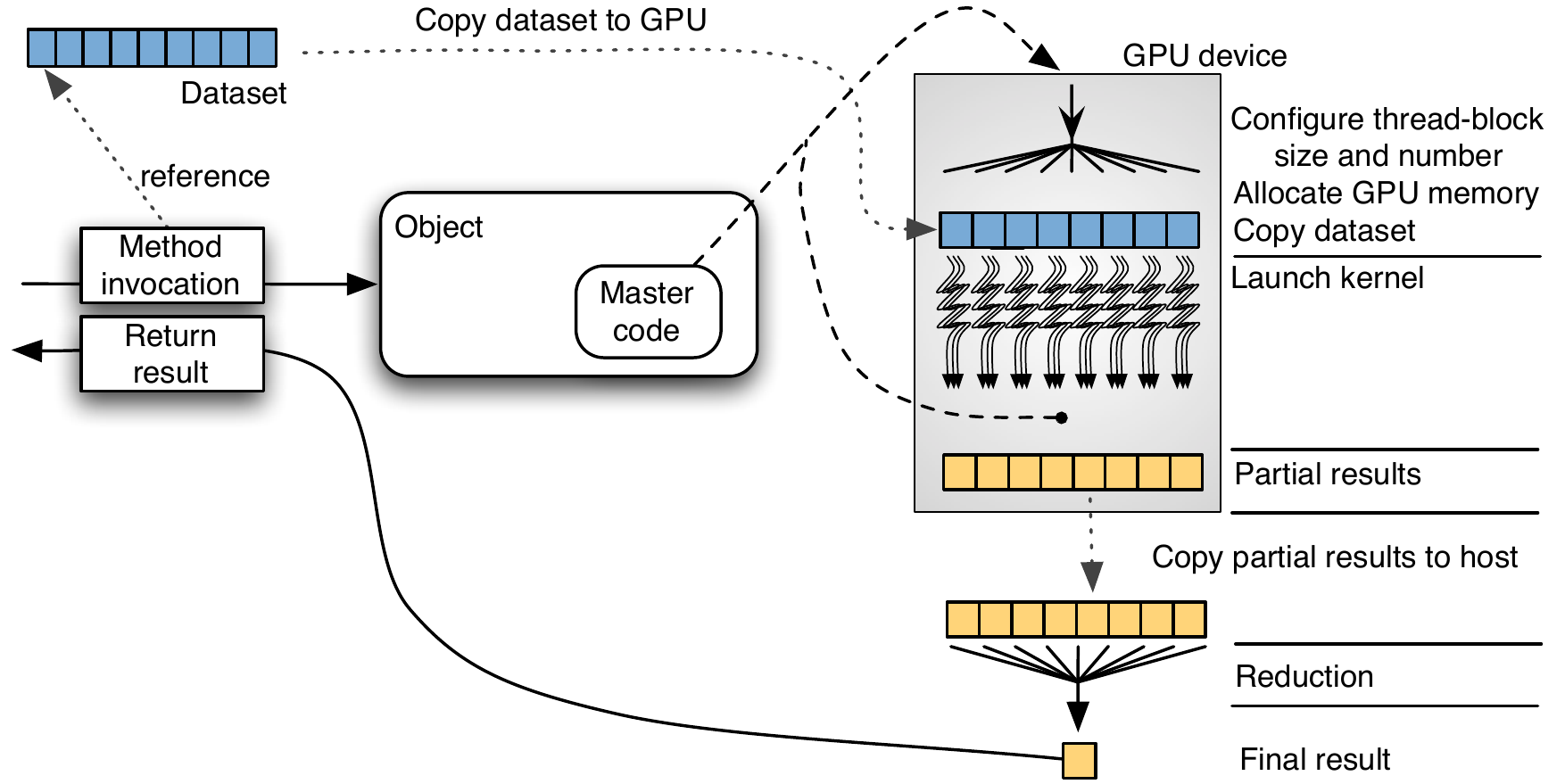}
\caption{SOMD execution model on GPUs}
\label{fig:somd-gpu}
\end{figure}

Given the number of thread-groups (work-groups in OpenCL) and the number of threads per group (work-items in OpenCL), 
domain decomposition is transparently  handled by the underlying execution model, both in OpenCL and in CUDA.
Therefore, the master code must only determine and supply such configuration values.
However, studies such as \cite{maestro} indicate that, 
if on one hand
the  determination of the number of thread-groups and their size is a crucial aspect for the computation's performance, on the other, the
 optimal solution is
 greatly dependent on the problem and on the target device(s).
Benchmarking is a common procedure for obtaining empirical performance information of the target devices and, with that,
adapt the work-loads to the devices capabilities.
In addition, statistics of the kernels' execution can also be taken into account.

As will be detailed in Subsection \ref{sec:comp:gpu}, the global MI synchronization induced by the \synck construct
forces  the code generated for the slave side  to  require more than one kernel.
Figure \ref{fig:somd-gpu} depicts the instantiation of the SOMD execution model in GPUs accordingly, a MI may comprise multiple kernel executions (of the same or possibly different kernels).
These kernels may  operate over different parts of the dataset, which  motivates an \textit{on-demand} copying strategy that makes use of the ability of current GPUs to overlap data transfer operations with computation.
Upon the final kernel execution, the partial results must be copied from GPU to host memory, for a complete or final reduction  (more details in Section \ref{sec:comp:gpu}) and posterior delivery of the computed result to the invoker.

\section{Compilation for Shared Memory and GPUs}
\label{sec:comp}

\begin{figure}
\centering
\includegraphics[width=10cm]{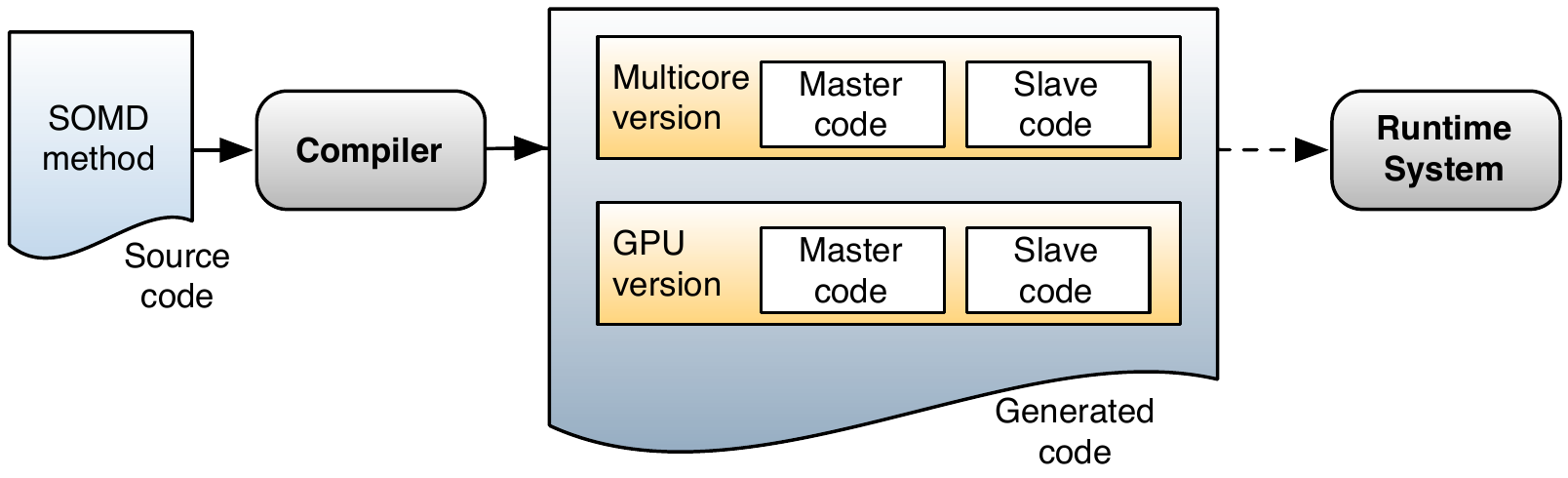}
\caption{Complier support for multi-architecture adaptation}
\label{fig:compiler}
\end{figure}

This section elaborates on how SOMD methods that operate upon arrays can be compiled for  multi-cores and GPUs.
To that extent,  the compiler  must produce two different versions of the code, one for each kind of architecture (Figure \ref{fig:compiler}).

The new constructs are translated into Java by a dedicated compiler built on top of 
   Polyglot  \cite{polyglot}, a  Java-to-Java compiler that provides a framework to easily extend the  language.
This compiler guarantees the program's type safety and generates Java code, which is in turn compiled by a standard Java compiler.

In order to engender the master-slave pattern, the compiler must generate code for both roles.
The master code  will replace the method's  original code and will have, as previously mentioned, the responsibility to launch and  coordinate the parallel execution of the MIs, and perform the reduction.
 The slave code will carry out a set of MIs upon one partition of the original dataset, on the target architecture.
While it is waiting, the master may itself execute one of the tasks. 
However, for  the sake of simplicity, here we have chosen to offload the entirety of the work to a pool of worker slaves.

We will now detail the code generation process for both target architectures.

\subsection{Shared-memory Architectures}
\label{sec:comp:sm}

The compilation process follows the guidelines of  Subsection \ref{sec:archs:sm}. The MIs are offloaded to a pool of threads, working in behalf of the application.
Each of these MIs will receive a range of indexes of the original array, that delimit the assigned partition. 
The outcome of their execution, the partial result, is placed on a position (given by the MIs rank)  of a results' vector created for that one purpose.
This array will be directly  fed to the reduction operation, which will take an array of elements of type $T$ and output a value of that same type.
Synchronization of task completion is performed through a phaser that we will refer to as \texttt{completed}.
The master blocks on this phaser, while the MIs simply notify  their completions and consequent writing on the results' vector.
Phasers are also used to encode the \synck construct, following the 
strict memory model premisses of Subsection  \ref{sec:archs:sm}. 
We will refer to this phaser as \texttt{fence}.

Algorithm \ref{algo:sm} sketches the compilation scheme for the master code.
The default computation of the index ranges over the annotated arrays is performed by  a dedicated partitioner (\texttt{IndexPartitioner}) at line 9.
The  partitioner  receives the length of the dimension to partition and the number of divisions, and returns an index range encoded in  a 2-element array.
These ranges are then supplied as argument to each MI (line 11). 

Data-structures that must be visible to all MIs, namely shared variables, the  \texttt{fence} phaser, and the \texttt{results} vector, are created by the master and passed  as arguments to the MIs (line 11).
The parameter list of the latter is completed with the value of their rank.

\begin{algorithm}[!t]
\caption{Compilation scheme for the master code}
\label{algo:sm}
\LinesNumbered
\begin{footnotesize}
\KwIn{SOMD method $m$ with parameters $L_{p}$ and return type $\tau$; a list of distributed parameters $L_d$;  a list of distributed local variables $L_l$; list of shared local variables $L_s$; and reduction $R$ with argument list $L_e$}
Declarations of initialized variables: \\
\Indp
  \texttt{\textit{nSlaves}}  $\leftarrow$ number of slaves assigned to the execution of the method \\
  \texttt{\textit{fence}} $\leftarrow$ phaser for encoding the \synck construct \\
   \texttt{\textit{completed}} $\leftarrow$ phaser to synchronize task completion \\
\Indm
\ForEach{s $\in L_s$}{
Declare $s$ and assign initial value (if any)
}
Declare the \texttt{\textit{results}} vector of type $\tau$ \\
\ForEach{a $\in L_d \cup L_l$}{
Generate statements to produce a set of \texttt{\textit{nSlaves}} index ranges of $a$ by resorting to partition strategy (\texttt{IndexPartitioner} by default)
}
\For{$i \leftarrow 0$ \KwTo $\texttt{nSlaves}-1$}{ 
 Generate  statements to create and spawn a task to enclose an MI working on the $i$th index range of each array $\in L_d \cup L_l$
}
Generate statements to: \\
 \Indp 
 Wait for the conclusion of the tasks in phaser \texttt{completed}\\
 Declare variable \texttt{result} of type $\tau$\\
Apply reduction $R(L_e)$ and assign result to variable \texttt{result}\\
Return \texttt{result}
\end{footnotesize}
\end{algorithm}

The code of a  slave task enclosing a MI results from a transformation of the method's original code.
We represent such transformation with function   \[\mathcal{C}: (SourceCode, Env)\ \mapsto\ SourceCode\]
where $Env$ is an environment that provides for the concrete identifiers of each of the compiler generated 
 parameters: \texttt{completed}, \texttt{fence}, \texttt{results}, \texttt{rank} and index ranges.
The translation function eliminates all declarations of shared variables, since these are included in the task's parameters.
The return of a result is translated into the writing of the value in the MI's index of the results' vector: \[\mathcal{C}(\textbf{\texttt{return}}\ e, \Sigma)  =  \Sigma(\texttt{results})[\Sigma(\texttt{rank})] \leftarrow e\ ;\ \Sigma(\texttt{completed}).\texttt{advance}()\]

The synchronization associated to  \synck blocks is
performed 
by a barrier at the end of the block's translation (by resorting to class \texttt{java.util.concurrent.\linebreak Phaser}): \[\mathcal{C}(\textbf{\synck} \{ P \},\Sigma)  =  \mathcal{C}(P, \Sigma)\   \Sigma(\texttt{fence}).\texttt{advanceAndWait}()\]

 The boundaries of  
\texttt{for} loops are modified to reflect the index range assigned to the MI.
 Consider a one-dimensional distributed array $a$ of size $N$.
Loop boundaries that encompass the entirety of the array ($[0, N[$) are translated into: \[[\Sigma(\texttt{a\_range})[0], \Sigma(\texttt{a\_range})[1][\]
Boundaries that only encompass a subset of the array's index space, such as $[e_1, e_2[$ with $e_1>0$ and $e_2 < N-1$
are translated into: \[[\texttt{max}(e_1,\Sigma(\texttt{a\_range})[0]), \texttt{min}(\Sigma(\texttt{a\_range})[1], e_2)[\]	
where $\texttt{max}()$ and $\texttt{min}()$ compute, respectively, the maximum and the minimum of the two given expressions.
When views are involved, these must be added and subtracted  to the upper and lower bounds, respectively.

Currently the loop transformations that we apply are quite simple and require the computation of the boundaries to be not dependent on local variables.
This restriction allows us to compute the number of elements to partition in the master code, and supply this information to the  index range partitioner.
Our future plans include applying polyhedral optimizations \cite{polyhedral} to enhance our loop parallelization. 

Listings \ref{lst:sm} and \ref{lst:sm-slave} depict, respectively, the generated master and slave code     for the stencil example of Listing \ref{lst:sor}.
The application of the\linebreak \texttt{IndexPartitioner} strategy takes into account the view specified for each dimension in the last argument.
Regarding the slave code, specialized versions can be generated for the first and last ranked MIs and with that remove the \texttt{max()} and \texttt{min()} operations.

\begin{lstlisting}[float, caption=Stencil example - Master code for shared memory, label=lst:sm,  language=somd]
double[][] stencil(double omega, double G[][], int num_iterations) {
 int nSlaves = getNumberOfWorkers(); // number of MIs
 Phaser fence = new Phaser(nSlaves); // Phaser "fence"
 Phaser completed = new Phaser(nSlaves+1); // Phaser "completed"
 double[][][] results = new double[nSlaves][][];  // "Results" vector
 int[][] G_1 = Distributions.IndexPartitioner((G.length-1), nSlaves, {1,1}); //	 1st dimension of G  
 int[][] G_2 = Distributions.IndexPartitioner((G[0].length-1), nSlaves,{1,1}); // 2nd dimension of G
 for (int i = 0; i < nSlaves; i++) // Spawn tasks
  spawn(new Stencil_MultiCore(omega, G, num_iterations, G_1[i], G_2[i], fence, completed, results, i));
 completed.advanceAndWait();  // Wait for their completion
 double result = Reductions.ArraySum(results); // Reduce results
 return result;  // Return result
}
\end{lstlisting}

\subsection{Graphic Processing Units}
\label{sec:comp:gpu}

The code generated for the master side follows the overall behaviour described in Subsection \ref{sec:archs:gpu}.
It must configure the number and size of the thread-groups, perform the required data transfers between the host and the GPU addressing spaces, launch the kernels, 
 perform the reduction stage, and return the result.
Other than specifying the thread grid, the master has no control on the data partition strategy.
The underlying GPU execution model does not permit such stage to be user-defined.
This limitation transits to the programmer, which means
that any partitioning strategy other than the default is ignored when generating the GPU code (a warning is emitted). 

The disjointness of the host and device addressing spaces
forces the master to 
allocate GPU memory for the  method's parameters and
  shared  variables, and also for the  \texttt{results} vector. 
Moreover, the initial contents of the parameters and the 
value assigned at shared variable declaration time (if any) must be copied
to the newly allocated positions.
The actual kernel execution is performed synchronously,  hence there is no need  for the \texttt{completed} phaser.
Once that last kernel terminates the 
 \texttt{results} vector is copied back to host memory, so it can be passed to the reduction stage.

\begin{lstlisting}[float, caption=Stencil example - MI code for shared memory, label=lst:sm-slave, language=somd, language=java]
class Stencil_MultiCore extends SOMDTask {
 // Method's original local parameters

 // SOMD generated parameters
 private final int[] G_1; 
 private final int[] G_2; 
 private final Phaser fence	;
 private final Phaser completed;
 private final double results;
 private final int rank;
 
Stencil_MultiCore(double omega, double[][] G, int num_iterations, int[] G_1, int[] G_2, Phaser fence,  Phaser completed, double results, int rank) {
    ... // assign parameters to local variables
 }
 
 void call() {
  ...
  for (int p = 0; p < num_iterations; p++) {
   for (int i = Math.max(1,G_1[0]); i < Math.min(G.length-1, G_1[1]); i++) 
    for (int j = Math.max(1,G_2[0]); j < Math.min(G[0].length-1, G_2[1]); j++) 
      // code of loop, line 8 of Listing 13
   fence.advanceAndWait();     
  }
  // Summation loop, lines 11 to 13 of Listing 13 
  results[rank] = Gtotal;
  completed.advance();
}
\end{lstlisting}

\begin{algorithm}
\caption{Compilation scheme for the master code in GPU}
\label{algo:gpu}
\LinesNumbered
\begin{scriptsize}
\KwIn{SOMD method $m$ with parameters $L_{p}$ and return type $\tau$; a list of shared variables $L_{s}$; a list of kernels $L_{k}$ resulting from the compilation of $m$; and reduction $R$ with argument list $L_e$}
Generate statements to determine thread-group number and size \\
\ForEach{a $\in L_p \cup L_s$}{
 Generate statements to allocate GPU memory for $a$ and copy $a$ to this newly allocated memory
}
Declare local copy of \texttt{\textit{results}} vector of type $\tau$ \\
Generate statements to allocate GPU memory for \texttt{\textit{results}} vector  \\
\ForEach{k $\in L_k$}{
 Generate statements to:\\
  \Indp 
  Synchronously launch k - parameters are all a $\in L_p \cup L_s$ and the \texttt{\textit{results}} vector  \\
  Eventually reduce some of the results on the host side \\
    \Indm 
}
Generate statements to: \\	
 \Indp 
 Copy contents of the \texttt{\textit{results}} to host memory\\
 Declare variable \texttt{result} of type $\tau$\\
Apply reduction $R(L_e)$ and assign result to variable \texttt{result}\\
Return \texttt{result}\\
\end{scriptsize}
\end{algorithm}

{\texttt{for} loops iterating over \distk annotated arrays must be refactored to generate a kernel that computes a range (usually of size one) of 
the iteration set.
A major concern in this refactoring is the memory access pattern.
Aligned memory accesses and thread convergence -  all threads participate in the memory  operations - 
are two very notable performance boosters in GPU programming.
Most loops in data-parallel algorithms iterate the entirety of one or more arrays, 
 thus access may be aligned and divergence is low.
 When multiple loops are involved, there is a high  probability of
 having  disparate numbers of iterations.
 This constitutes a problem, because the number of threads required to execute the refactored code of each loop is different.
 We have therefore to find a trade-off between aligned accesses and non-divergence.
We favour the first, since their impact of performance is generally lower.
Given this, the translation of a loop is given by:
\begin{eqnarray*}
\mathcal{C}(\textbf{\texttt{for}}_i (e_1; e_2; e_3)\ \{\ P\ \}, \Sigma)\  & = & \textbf{int}\ \Sigma(\iota_i)  = \texttt{getGlobalId()}\ ;\\
 &&\textbf{if}\ (\Sigma(\iota_i) >= lb_i\ \textbf{and}\ \Sigma(\iota_i) < ub_i)\ \{ \mathcal{C}(P, \Sigma) \} 
\end{eqnarray*}
where
$i$ denotes the rank of a loop within a  SOMD method, according to a total order established from the line number by which loops appear in the source code, $lb_i$ and $ub_i$ are the loop's lower and upper bounds, and $\iota_i$ the  loop's induction variable, obtained from a previous  analysis of the code.
The concrete transformation of the 
first loop of the stencil example is illustrated in Listing \ref{lst:loop}.  

\begin{lstlisting} [float,language=somd,caption=Transformation of the 
first loop of the stencil example, label=lst:loop]
int ij = getGlobalId();
int i = ij/size; // Obtaining the i index from the one-dimensional grid
int j = ij%size; // Obtaining the j index from the one-dimensional grid
if (i>=1 && i < size-1)  // Checking loop boundaries
 if (j>=1 && j < size-1) // Checking loop boundaries
  G[ij] = (G[(i-1)*size+j] + G[(i+1)*size+j] + G[ij-1] +  G[ij+1]) +
               a_constant * G[ij]; // Calculus, the matrix must be flattened
\end{lstlisting}

There is divergence on the boundary groups, since  one or two threads (depending on the group's position in the grid) will not perform the computation.
In fact, there are many factors that establish the overall performance of the kernel, such as the use of local memory, eliminating bank conflicts, loop unrolling, and so on.
To prototype our solution in Java we resorted do AMD's Aparapi API \cite{aparapi} and, hence, delegated most of these concerns on a lower-layer of our software stack.
However, the use of Aparapi raises other issues, of which the more limiting are the lack of support for arrays with more than one dimension, and of double precision arithmetic. 
The first limitation is visible in the flattening of the array in line 6 of Listing \ref{lst:loop}.
Rootbeer \cite{rootbeer} could be a viable alternative,  since it allows more of Java's features to be used in the programming of a kernel.
However, its runtime makes use of a dedicated GPU memory manager that does not allow memory objects to reside in the GPU's memory across multiple kernel executions. 
This feature is a requisite of our compilation process,  as will be detailed in the remainder of this subsection.

Loops that perform reduction operations within the method's body, such as the second loop of the stencil example, also require special attention.
In GPU devices, these reductions can be performed  globally by all threads.
However, GPUs are not particularly efficient when reducing a full data-set into a single scalar value. Instead, it is preferable 
to begin the enterprise on the device, and 
move it to the host side as soon as there is not enough work to keep all of the GPU's processing units busy, a threshold that  is device-dependent.
This approach justifies line 9 of Algorithm \ref{algo:gpu}.

When a reduction is applied at the end of a kernel, it is most likely to be the same as the one provided in the \reducek construct.
When such is the case, only the latter is performed.
In an opposite case, when the method does not end with a reduction,
this overall reduction compilation strategy can  be used to parallelize the reduction provided in the \reducek construct.

\paragraph{Configuration of the Thread Grid}
The determination of the  thread-group number and size (line 1 in Algorithm \ref{algo:gpu}) 
adjusts the total number of threads according to the maximum size allowed for a thread-group 
(local work-size in Aparapi and OpenCL) in the target device.
For instance, if such value is 512, and the size of the problem equals 1000000
\[\texttt{numberOfThreads}(1000000) = 1 000 448 = 1954 \times 512\] resulting in 1954 groups of 512 threads. Naturally, some of these will not perform any effective computation, since they fall outside the loops' boundaries.

\paragraph{Data Dependencies and Synchronization}
Global, inter-group, synchronization is not permitted in the GPGPU execution model.
Group independence is an established pre-condition that permits an efficient, hardware implemented, scheduling of the computation.%

Consequently, the \synck data-driven synchronization construct
cannot be trivially translated  into either OpenCL, CUDA, or any other GPGPU framework.
The only global synchronization point is implicitly established 
when the control is relinquished back to the host, upon kernel completion.
Given this, the compilation of \synck  
requires the synchronous iterative issuing of a kernel
responsible for executing the code enclosed by the construct.
Coming back to the stencil example, Listing \ref{lst:master-aparapi} highlights a snippet of the code generated for the master, resorting to the Aparapi API.
Line 1 showcases the determination of the total  number of GPU threads, whilst \texttt{kernel}, declared at line 2, refers to the code enclosed by \synck.

\begin{lstlisting}[float, caption=Snippet of the code generated for the master, label=lst:master-aparapi, language=somd]
int nThreads = numberOfThreads(G.length*G[0].length); // Size of the matrix, for readability's sake
Kernel kernel = new Kernel1(G);
kernel.setExplicit(true); // Explicit management of the data transfers
kernel.put(G); // Transfer the G matrix
for (int p = 0; p < num_iterations; p++) 
 kernel.execute(nThreads);  // Execute the loop - lines 6 to 8 of Listing 13
kernel.get(G);  // Read the matrix to the host
\end{lstlisting}

\section{Runtime System}
\label{sec:java}
 
 The duo compiler/runtime system  that backed the results presented in \cite{somd-hpcc} resorted to  X10's Java runtime system (X10RT).
Since then  we shifted to the Elina parallel computing framework \cite{elina}.
Elina is highly modular, though perfect for the development of 
new parallel runtime systems, since it allows for rapid prototyping and, posterior refinement and optimizations.
Moreover it suits perfectly our vision of code once, run in multiple architectures.

One of the responsibilities delegated onto the runtime system is the choice, for each SOMD method,
of which version to execute, among the multiple versions generated by the compiler.
 This choice must take into consideration the characteristics of the underlying architecture, in particular its ability to execute the selected version,
and the configuration information provided by the user.
Concerning the scope of this paper, stand-alone computers, the shared memory version is selected by default.
The user may force GPU execution by providing a configuration file composed of rules of the 
form: \texttt{Class.method:target\_architecture}.
The inapplicability of the user's preferences, given the available hardware, reverts to the default setting.

Some of the master's code generated by Algorithms \ref{algo:sm} and  \ref{algo:gpu}, is factorized by Elina.
Among these is the spawning of the multiple tasks, and the application of the reduction function. 
These factorizations have not been exposed in the algorithms with the purpose of providing a higher degree of abstraction, removed from  overwhelming details of the  Elina  API.

Regarding the parallel computing engine, the shared memory support resorts to a Java thread-pool, parametrized and managed by Elina.
The default parametrization takes into account the number of cores available in the system, but this 
 setting may be overridden  both at development and/or deployment time.
 SOMD execution requests may be submitted concurrently, and hence compete for this pool of threads.
 The scheduling and load balancing is internally managed by the system.

The current GPU support is experimental. We delegate mostly everything to the Aparapi library, which, in turn, resorts to OpenCL.
Our objective is to have such support seamlessly integrated in Elina, but we are still waiting for a  Java GPGPU API that meets our requirements.

	\section{Evaluation}
\label{sec:eval}

This section evaluates our approach from a functional and performance perspective.
For comparison purposes, we  implemented a subset of the JavaGrande (JG) benchmark suite \cite{javagrande},  namely the applications of its Section 2 for which the suite provides multi-threaded implementations.
We built from  the sequential implementations  of these same benchmarks (also included in the suite), being that our intervention
was limited to the annotation of the code as it was. No extra optimizations were performed.
Moreover, we tried to resort, as much as possible, to built-in features. 

In this section we begin by briefly describing our take on the implementation of each of these applications, and evaluate the impact on the original sequential code. Next we carry out a comparative performance analysis in both shared memory and GPUs.

\subsection{Benchmarks}

\paragraph{Crypt} Ciphers and deciphers a given sequence of bytes.
We implemented each of these operations as a  SOMD method that, given    the original 
byte array, returns its cipher. 
We qualified both original and destination (allocated within the method's scope) arrays
with \distk, applying the built-in array partitioning strategy.
The 	method's body comprises a single loop that traverses the entirety of both  arrays, unrolled so that each iteration operates upon eight bytes.

\paragraph{LU matrix factorization (LUFact)}  Linear system solver that uses LU factorisation followed by a triangular solve. 
The benchmark only parallelizes the  factorisation stage.
The JavaGrande multi-threaded version does so by using a ranking scheme to divert the kernels.
The prime (with rank 0) is the sole to perform all operations. The remainder only enter the scene to perform a loop of invocations to an implementation of the \textit{Daxpy} BLAS routine.
This requires a number of synchronization points, namely 6 barriers, although some could be eliminated.

Our proposed programming model is rank agnostic. Therefore our approach was to decompose the algorithm into two methods.
The top-level one performs the main iterative loop and resorts to an \textit{actual}
 SOMD method to apply parallelism where needed.
 Hence, since the execution of a SOMD method is synchronous, no explicit synchronization points are required.

\paragraph{Series} Computes the first $N$ Fourier coefficients ($N$ given as argument) 
 in interval [0,2].
The method is parametrized in a single argument, a matrix with two rows - one for the $a_n$ terms and another for the $b_m$ terms, for
  $n \geq 0$ and $m \geq 1$.
  In JavaGrande's implementation, the computation of $a_0$ is performed by a single thread, selected by its rank.
As in the LUFact case, our solution resorts to two methods, the top-level one simply computes $a_0$ and invokes a SOMD method
to perform the rest of the job in parallel.
Since the input matrix only features two rows, only the column dimension  is partitioned: \distk(dim=2).

\paragraph{SOR}  Solves a system of linear equations of size $N \times N$ 
through Jacobi's \textit{Successive Over-Relaxation} numerical algorithm.
 $N$ is given as argument, while the number of iterations is set at 100.
As illustrated in Listing \ref{lst:sor}, the input matrix is partitioned through the built-in strategy, that performs an index range
 distribution on both dimensions - the equivalent to a (block,block) distribution.
The method's body features a single loop that requires a \synck block.

\paragraph{Sparse Matrix Multiplication (SparseMatMult)} Performs a multiplication over a matrix	of size
$N \times N$ in compressed-row format.
The vectors with the matrix's data, row index and column index are all partitioned through a user-defined strategy that 
ensures the disjointness of the ranges of rows assigned to each partition. 
The method's body is straightforward and thus required no additional annotations.
The user-defined distribution applies the  algorithm 	
featured in  JavaGrande's multi-threaded version of the benchmark, which contains roughly 50 lines of code.

\subsection{Performance Analysis (Shared Memory)}
This analysis  compares the   performance of our implementations against the  multi-threaded versions  featured in the JavaGrande suite.
 The measurements account  the  parallel  decomposition of the problem plus the actual 
  execution of the computational kernel.
 The presented speed-up values  are relative to the original sequential versions,
 and result from an average of the middle tier of 30 measurements.

\subsubsection{Infrastructure and Benchmark Configurations}
  All measurements were performed on  a system composed of two Quad-Core AMD Opteron 2376 CPUs 
 at 2.3 GHz and 16 Gigabytes of main memory,  running the 2.6.26-2 version of the Linux operating system. 

The JavaGrande suite 	specifies three configuration classes (A to C) for each benchmark.
Table \ref{tab:classes} presents such configurations, as well as our baseline - 
the execution times of the original sequential implementations.

\begin{table}
\centering
\caption{Reference table for the benchmark's configuration classes}
\label{tab:classes}
\begin{scriptsize}
\begin{tabular}{|l||c|r|}
\hline
 Benchmark & \multicolumn{1}{c|}{Configuration} & Execution time ($s$) \\
 \hline
\hline
  \multicolumn{3}{|c|}{Class A} \\
\hline
\hline
\textit{Crypt} & vector size:  3000000   &   0,225 \\
\hline
 \textit{LUFact} & matrix size:  500   &  0,091 \\
\hline
 \textit{Series} & Number of coefficients: 10000  & 10,054\\
\hline
  \textit{SOR}  & matrix size:  1000  & 0,885 \\ 
\hline
\textit{SparseMatMult} & matrix size:  50000 & 0,665 \\
\hline
\hline
 \multicolumn{3}{|c|}{Class B} \\
\hline
\hline
\textit{Crypt} & vector size:  20000000   &   1,341 \\
\hline
 \textit{LUFact} & matrix size:  1000   &  0,778 \\
\hline
 \textit{Series} & Number of coefficients: 100000  & 102,973\\
\hline
  \textit{SOR}  & matrix size:  1500  & 2,021 \\ 
\hline
\textit{SparseMatMult} & matrix size:  100000 & 1,744 \\
\hline
\hline
 \multicolumn{3}{|c|}{Class C} \\
\hline
\hline
\textit{Crypt} & vector size:  50000000   &   3,340 \\
\hline
 \textit{LUFact} & matrix size:  2000   &  9,181 \\
\hline
 \textit{Series} & Number of coefficients: 1000000  & 1669,133\\
\hline
  \textit{SOR}  & matrix size:  2000  & 3,432 \\ 
\hline
\textit{SparseMatMult} & matrix size:  500000 & 19,448 \\
\hline
\hline
\end{tabular}
\end{scriptsize}
\end{table}

\subsubsection{Results}
The charts of Figure \ref{fig:gr1} to \ref{fig:gr3}
depict the  speed-up results obtained by both implementations,  for each class of configurations.
The number of partitions (in the SOMD versions) or threads (in the JavaGrande versions)
range from 1 to 8. The first allows us to assess the overhead imposed by either approach, while the second explores the maximum parallelism available.

\begin{figure}[h]
\centering
\begin{subfigure}{0.48\textwidth}
\includegraphics[width=\linewidth]{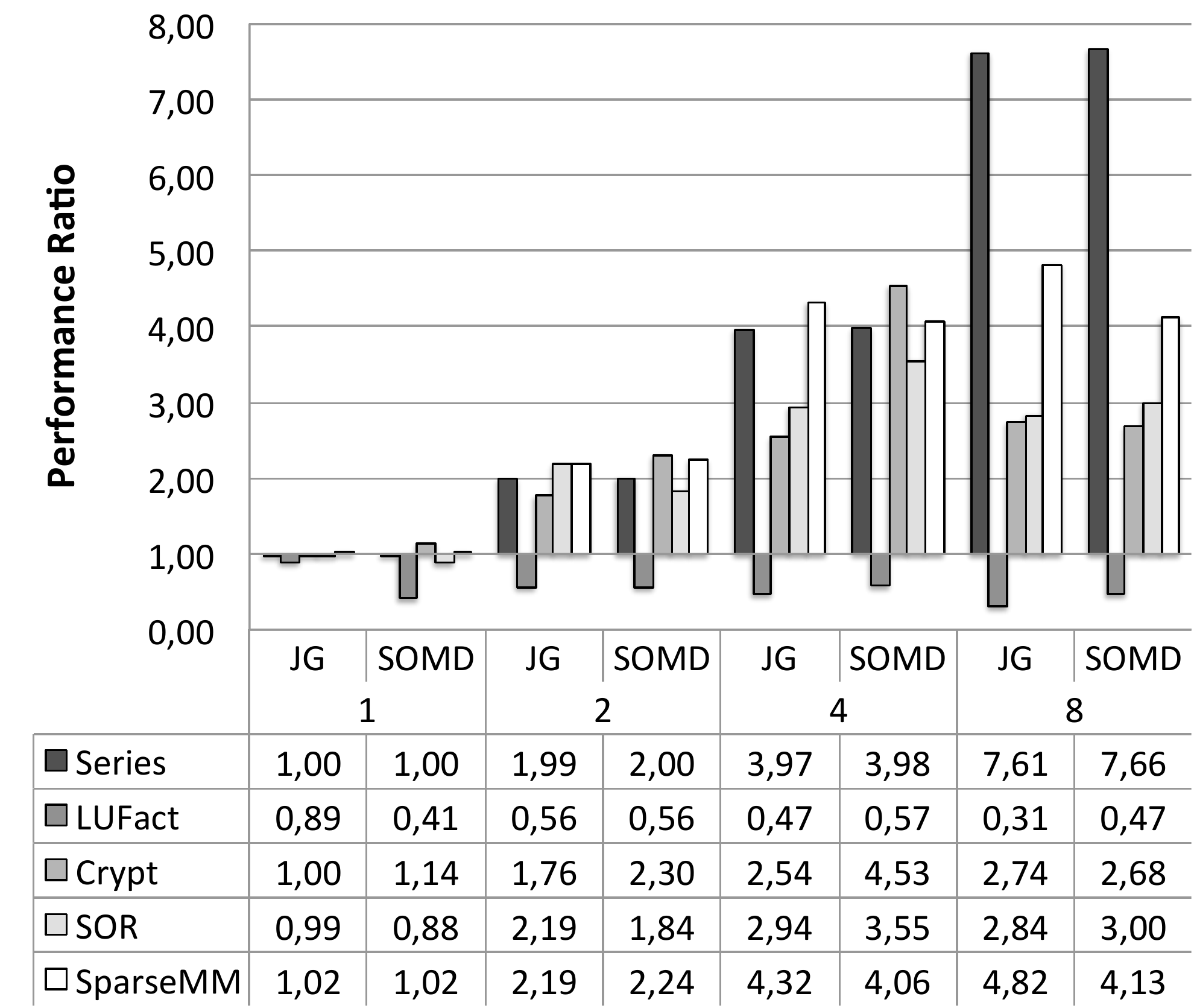}
\caption{Class A}
\label{fig:gr1}
\end{subfigure} \quad
\begin{subfigure}{0.48\textwidth}
\centering
\includegraphics[width=\linewidth]{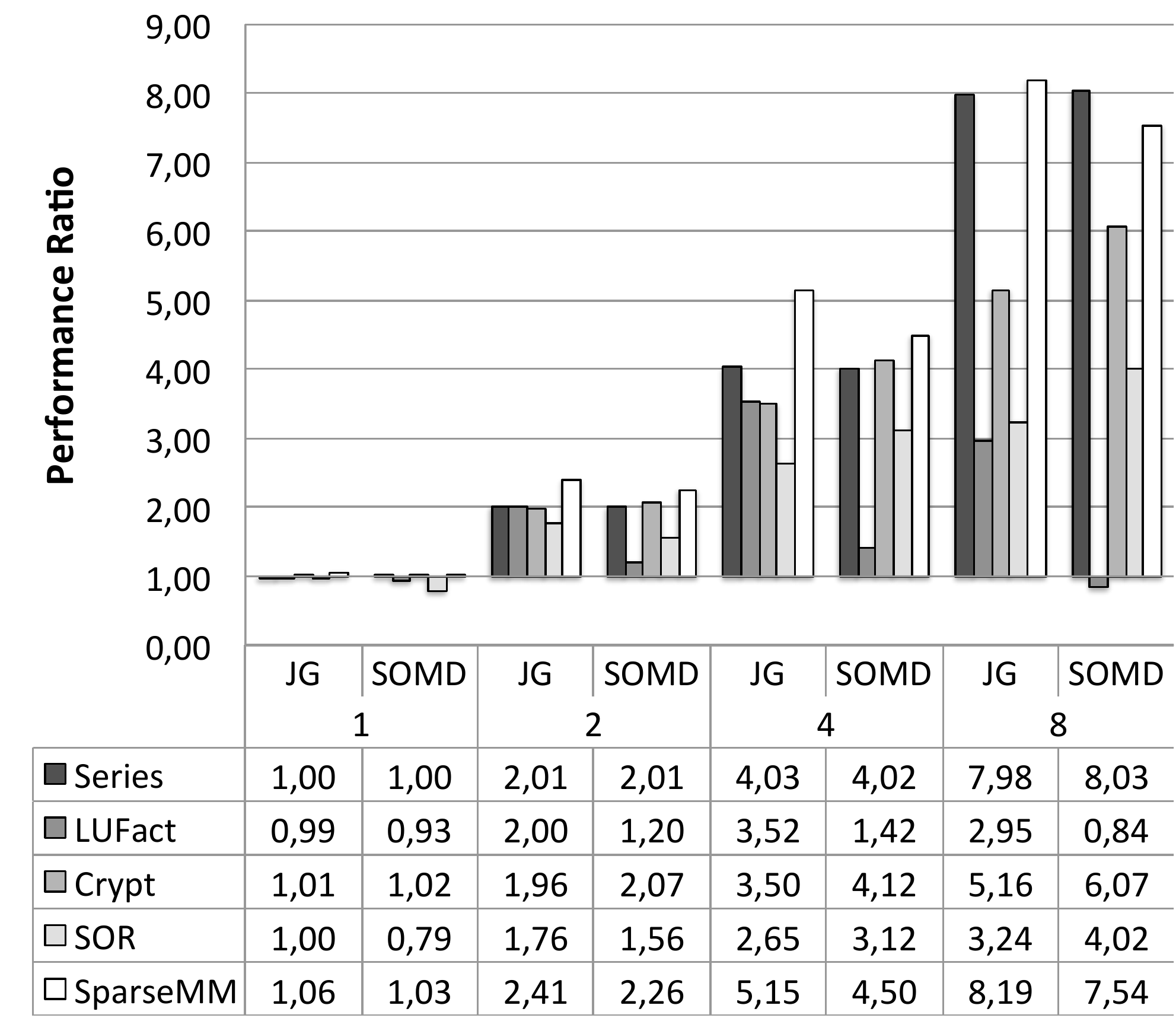}
\caption{Class B}
\label{fig:gr2}
\end{subfigure}
\begin{subfigure}{0.48\textwidth}
\centering
\includegraphics[width=\linewidth]{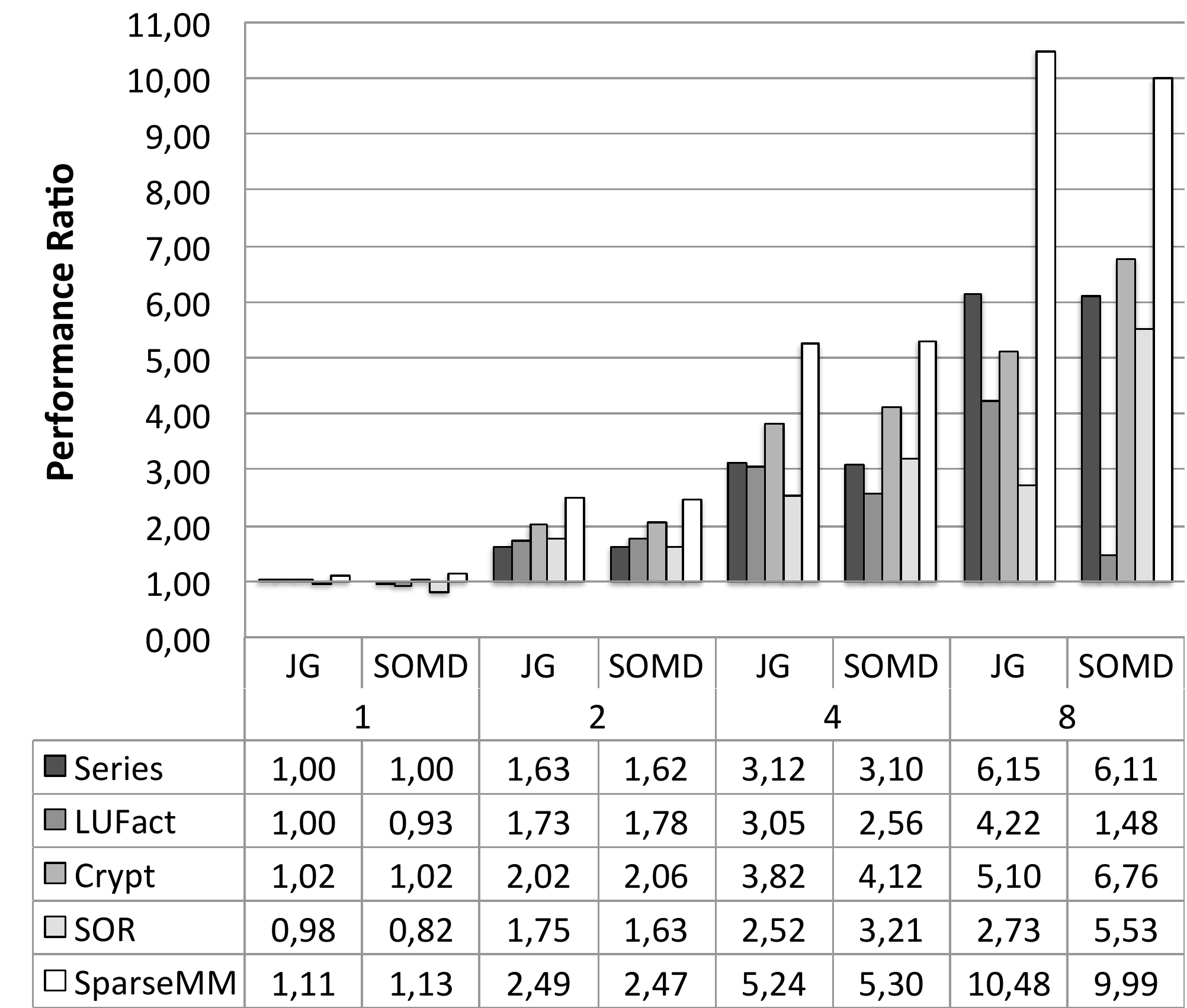}
\caption{Class C}
\label{fig:gr3}
\end{subfigure}
\caption{Speed-ups for shared memory relatively to the JG sequential version}
\end{figure}

\paragraph{Crypt} The SOMD approach scales better in all configurations.
The main reason for this performance delta is our optimized index range partitioning algorithm, which proves to be considerable faster than
JavaGrande's. 
The performance are only on par when 
the weight of the kernel is too small for these optimizations to have impact (Class A, 8 partitions).

 \paragraph{Series} The benchmark's long execution times mitigate the overhead imposed, and the optimizations performed, by any of the approaches.
 Ergo, the results are on a par in all classes.

\paragraph{SOR} Once again our partitioning algorithm makes the difference.
JavaGrande's version only parallelizes the outer loop, meaning that each thread receives a range of rows.
Our built-in approach performs a (block, block) distribution, generating a list of smaller matrices that allow for both loops to be parallelized.
In practice, we generate a more cache friendly code that takes advantage of both spatial and temporal locality.
This is reflected in the results, our work distribution may be heavier (we loose in the 2 partition configuration in all classes) but provides much better results as the problem size increases.

 \paragraph{SparseMatMult} 
 The only benchmark to require a user-defined partitioning strategy.
This strategy was borrowed from  the JavaGrande version, thus
 no performance penalty should incur from it. Given that the computational kernel is almost 
 exactly the same, the reasons behind JavaGrande's overall best performances must be in the overhead imposed by the Elina runtime system.

 \paragraph{LUFact}  We saved  for last the benchmark for which the proposed  programming model was not able to produce good results. 
  JavaGrande resorts to a rank-based approach to distinguish one of the threads (with rank 0) from the remainder.
  Accordingly, it may distribute the work and launch the required threads only once, at the beginning.
  Any sequential part of the algorithm  is executed only by this ranked-0 thread.
 This is done at the expense of having to explicitly synchronize the execution of the threads.
Our approach, although more declarative and absent of explicit synchronization, follows a split-join pattern.
Each invocation to the inner SOMD method requires the partitioning of the input dataset and the spawning of  tasks.
This is a viable solution for workloads heavy enough to dilute this extra overhead, which is not the case of LUFact.
 The execution time of the section suitable for parallelism is 0.010, 0.013, 0.021 for the A, B and C configuration classes, respectively. 
Consequently, the results were below par.
No implementation was able to deliver speed-ups for Class A, but, as the problem size increased, the JavaGrande 
version came out with the best performances. 
The SOMD version was able to even things up on Class C, while the weight of the parallel section so allowed.

\subsection{Performance Analysis (GPU)}
Our next analysis takes, for each configuration class,  the best results delivered by the  versions evaluated in the previous subsection, and compares them against the GPU code generated from the same SOMD source.

\subsubsection{Infrastructure}   
  The measurements were performed on  two GPU-accelerated systems:
\begin{itemize}
\item the first, that we will name \textit{Fermi},  is composed by a Quad-core Intel Xeon E5506  at 2.13 GHz, 
 and a NVIDIA Tesla C2050 GPU with 3 Gigabytes of memory.
The operating system is Linux, kernel version 2.6.32-41.

\item the second, that we will name \textit{GeForce 320M}, is an Apple MacBook Pro Laptop featuring an Intel Core 2 Duo processor at 2.4 GHz,
 and a NVIDIA Geforce 320M GPU that as assigned 256 Megabytes of shared memory.
\end{itemize}

\subsubsection{Results}
The results are displayed in charts \ref{fig:gr4} to \ref{fig:gr6}.
We have omitted LUFact, since the limitations that bounded the performance results in the shared memory infrastructure,
were inflated in the GPU. Each invocation of the SOMD method requires copying the whole matrix to the GPU, since it is modified between invocations,
 and launching the execution of a  kernel  that is not computationally heavy enough to mitigate the overhead of the data transfers.

\begin{figure}[h]
\centering
\begin{subfigure}{0.45\textwidth}
\includegraphics[width=\linewidth]{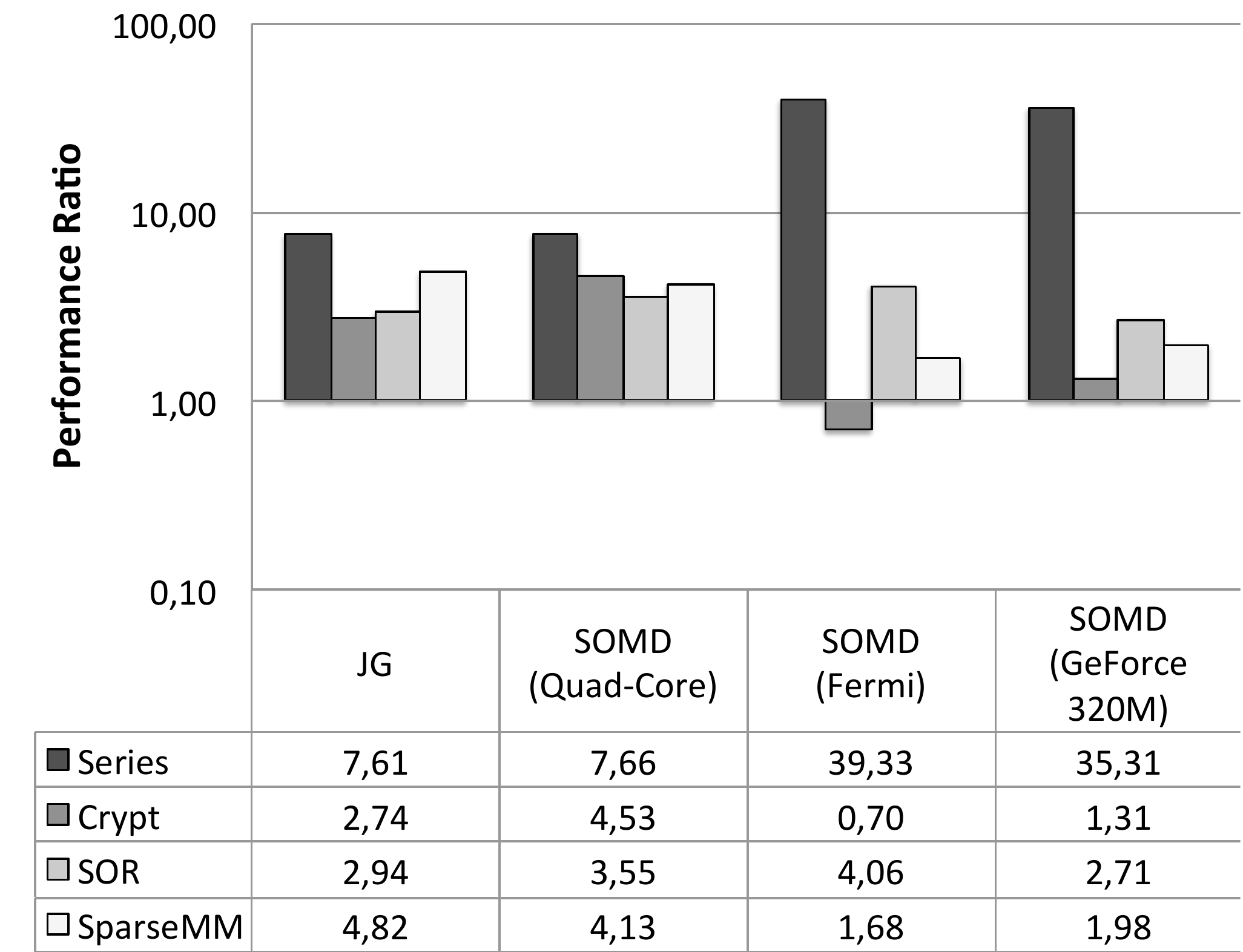}
\caption{Class A}
\label{fig:gr4}
\end{subfigure} \quad
\begin{subfigure}{0.45\textwidth}
\centering
\includegraphics[width=\linewidth]{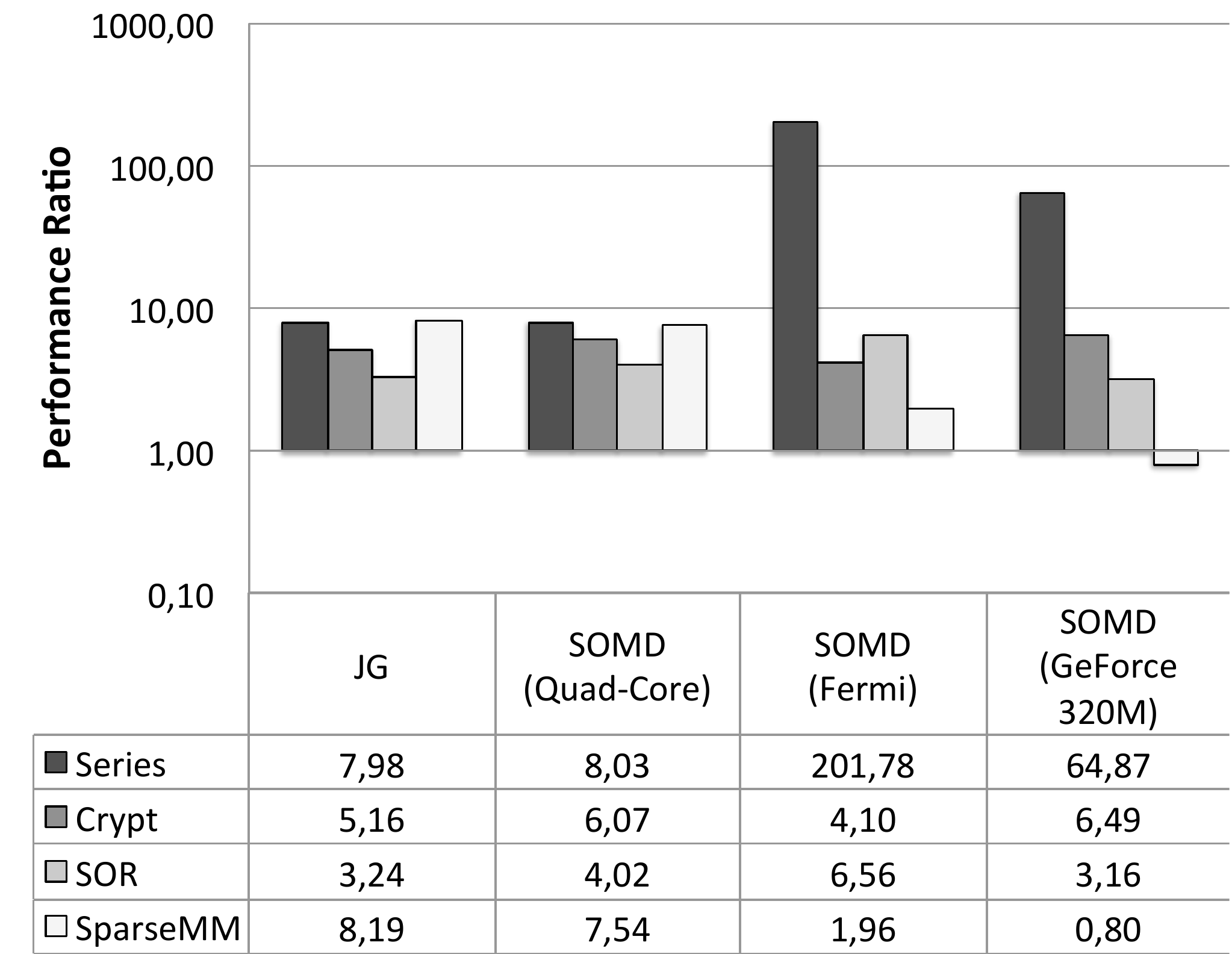}
\caption{Class B}
\label{fig:gr5}
\end{subfigure}
\begin{subfigure}{0.45\textwidth}
\centering
\includegraphics[width=\linewidth]{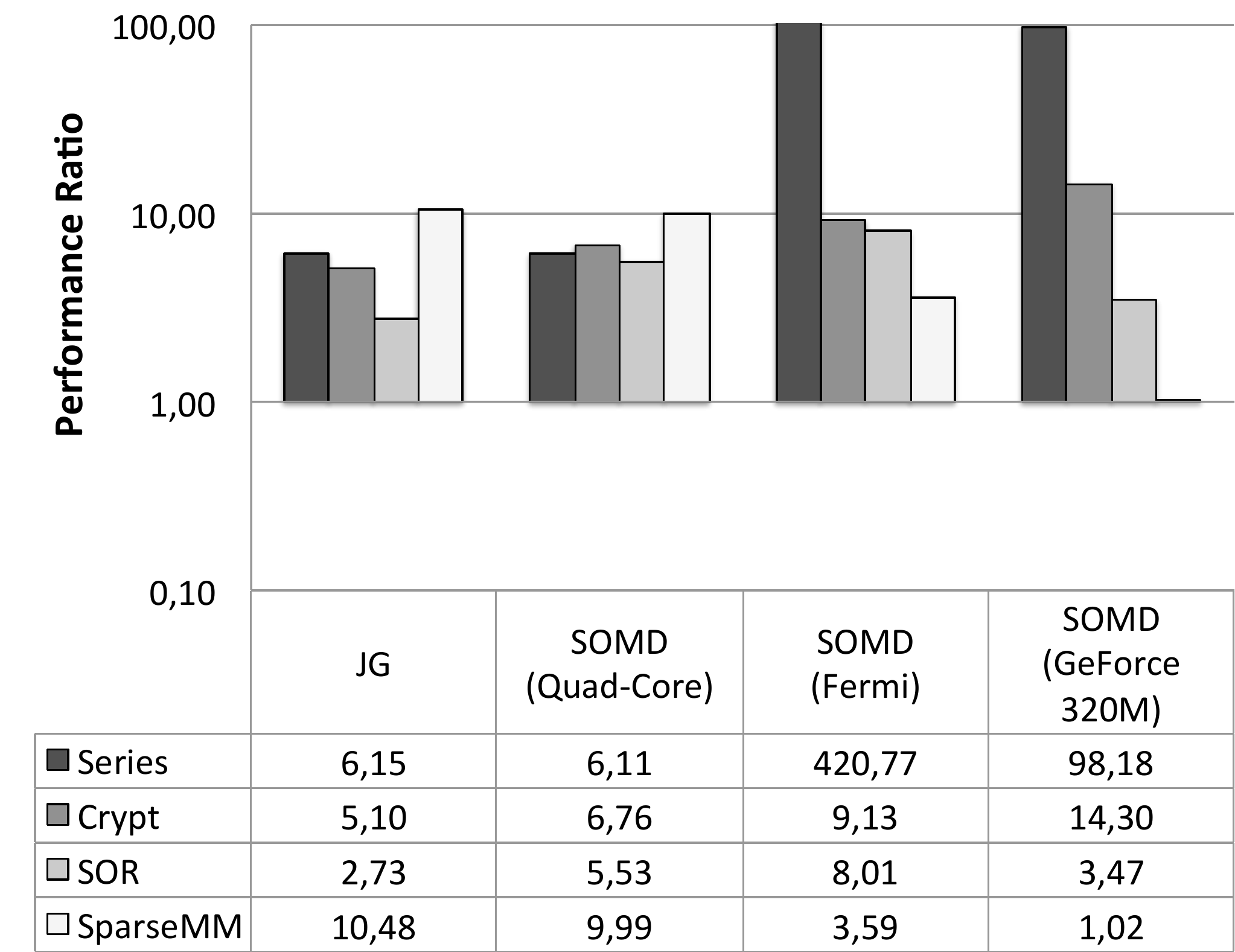}
\caption{Class C}
\label{fig:gr6}
\end{subfigure}
\caption{Best JG  and SOMD CPU versions versus
the GPU SOMD version. The depicted speed-ups  
are relatively to the JG sequential version}
\end{figure}

\paragraph{Crypt} is a memory-bound benchmark in this setting. The computational kernel is too small to justify the overhead of copying the data to the GPU.
Therefore, the shared memory versions deliver the best results for classes A and B.
The impact of the data movement overhead is also noticeable in class C; by sharing memory with the CPU, the GeForce 320M outperforms the Fermi, despite having  much less computational power.

\paragraph{Series} greatly benefits from GPU execution, as it is a heavy computing bound application.
 The speed-ups range from 39.46  to 420.77  in the Fermi infrastructure, and from 35.42 to 98.18 in the GeForce 320M.
Note that, due to Aparapi's inability to generate kernels that require double precision, we had restrict ourselves to single precision.
Ergo, the results are not as accurate as in the shared memory versions.
This also has a major impact in the execution time, since the throughput of double precision in, for instance, the Fermi GPU is, at its peak, half the one for single precision.
This is most noticeable in the transcendental and trigonometric functions that are massively exercised in this benchmark.
The GPU version of  benchmarks SOR and SparseMatMult also performs single precision, rather than double precision, arithmetic. 

\paragraph{SOR} is the only application in JavaGrande's section 2 whose SOMD version requires a \synck block.
Consequently, the generated GPU version features a loop on the host side that, at each iteration, prompts a kernel execution request. 
Aparapi's  explicit data movement management facilities allows us to transfer the input matrix only  once, at the beginning of the computation.
Nonetheless, the multiple, 100 to be exact, kernel execution requests convey a non-negligible overhead.
Even so, the best overall performances are achieved in the Fermi infrastructure.

\paragraph{SparseMatMult} exercises indirect memory accesses, which do not really fit in the GPGPU model, since they break the coalescing of memory accesses.
Moreover, the user-defined partitioning strategy is ignored (see Subsection \ref{sec:comp:gpu}).
Therefore, there is no guarantee that different thread-groups access different rows in the matrix, which may lead to  conflicts in  global memory accesses.
The results confirm these limitations, as none of GPU accelerated systems are able to be competitive with the shared memory versions.

\subsection{Discussion}
The evaluation attested the viability of the proposed model in the programming of both multi-core CPUs and GPUs.
The SOMD shared memory versions are competitive against the  multi-threaded versions provided by the JavaGrande benchmark suite, which perform  hand-tuned parallel decompositions of the domain across a given number of threads.
In fact, in some cases we were able to deliver better performances without extra programming support (other than the original method), due to our built-in partitioning strategies.

\begin{table}
\centering
\caption{SOMD adequacy of JavaGrande's section 2}
\label{tab:jg}
\begin{scriptsize}
\begin{tabular}{|l||r|r|}
\hline
Benchmark & Number of annotations & Extra  LoC \\
\hline
\hline
\textit{Crypt} & 2 & 1  \\
\hline
\textit{LUFact} & 1 & 3 \\
\hline
\textit{Series} & 1 & 3 \\
\hline
\textit{SOR} & 2 & 1  \\
\hline
\textit{SparseMatMult} & 3 & ~50  \\
\hline
\end{tabular}
\end{scriptsize}
\end{table}

Table \ref{tab:jg} presents the number of annotations, and extra lines of code
required by our implementations.
We do not take into consideration block delimiters in this count.
The impact on the code is minimal, which
corroborates our statement that the proposed model 
is simple to use.
This affirmation is further  substantiated with the facts that:
1) built-in partitions cover most of the needs when handling arrays; 
2) most of the user-defined partitioning  and reduction strategies are general enough to be available in a library, and therefore 
applicable in multiple scenarios, and
3)  the computations  performed by the latter   are  algorithmic problems that do not require special knowledge of parallel programming.

The GPU support gave us an  assessment of 
the expressiveness of our constructs in this particular context.
Naturally there are also some limitations that we will discuss ahead.
Nonetheless, we were able to easily offload the benchmarks to the GPU with good initial results.
We also noted that our annotation based approach is somewhat in-between the implicit and explicit memory transfer management 
support provided by Chapel, X10 and Aparapi.
The scope of the method determines  the boundaries of data transfers, being that this data persists on the GPU until the computation of the method, which may encompass several kernels, terminates.
Ergo, the method's boundaries implicitly  play   a role comparable to  \textit{data} regions in OpenACC \cite{openacc}.

These experiments on heterogeneous architectures also supplied us an initial assessment of
which kind of applications should be offloaded to the  GPU, and which should be executed  by the CPU.
We will build upon this work to try to infer these characteristics automatically, in order to provide an initial configuration that may be adjusted at runtime, as actual execution time information can be retrieved. 
These execution times may also be persisted, so that can be later employed  in the refinement of the configuration of future deployments.

\subsection{Limitations}

The SOMD model is specially tailored for data-parallel computations whose domain can be decomposed into independent partitions, and the 
results of the partial computations subsequently reduced.
Therefore, the performance of a SOMD method is tightly coupled to the number of  non-scalar parameters that are effectively distributed.
For instance, a SOMD take on matrix multiplication (MM) should resort to solutions that decompose both matrices, rather than only one. 
Moreover, data sharing may also introduce  overheads.			
These issues have more impact on some architectures than others: 
\begin{description}
\item[Multi-core CPUs]  are optimized for shared-memory parallelism.
Hence, the second matrix in MM could be shared by all MIs with apparently minor performance penalties.
However, assigning bigger datasets to each MI results in less cache friendly code, a crucial factor when performance optimizations are at stake.
The same reasoning can  be applied for the writing on shared, rather than private, data.

\item[On cluster environments] undistributed parameters increase the amount of data to be transferred  to each node. 
To a greater extent, the use of shared data infuses network communication and synchronization in MI execution, which are known to be performance bottlenecks.
In \cite{elina} we provide some initial experimental data on this topic.

\item[On GPUs] data sharing may cause bank conflicts, since multiple threads from distinct thread-groups  may try to concurrently access the same memory locations.
This is further aggravated by the fact that, on GPUs,  there  is no architectural support for global synchronization.
Thus, writing on data shared across all MIs may compromise the correctness of an algorithm. 
These scenarios can be statically identified and a warning issued.
\end{description}

The SOMD approach is also not particularly suitable when the data-parallel computation (the SOMD method) must be  iteratively applied,
as it is noticeable in the LUFact benchmark. 
The overhead of a per iteration distribution and reduction mines  the global performance.
This limitation can be easily overcome by extending the programming model with a  construct (\singlek) to delimit sections of the code that must be executed by a single MI. 
We are, however, trying to solve this issue under the wood, in the context of the compiler and the runtime system.
Our success will avoid having to expose these details to the programmer.

In conclusion, whenever performance is the most important requirement, SOMD application should be confined to 
data-parallel computations that can be decomposed into independent partitions.
This sometimes  implies  altering the original sequential implementation.
Otherwise, the annotations can be used directly upon the unaltered sequential code as the trade-off between performance and productivity.
In fact, our experience \cite{somd-hpcc} reveals that many of the operations performed on vectors and matrices can be trivially parallelized only by applying these annotations on the non-scalar parameters. In such cases, no trade-off is necessary.

Regarding our current GPU support, the  back-end is a proof-of-concept prototype, and naturally there is much room for improvement.
The use of Aparapi as an intermediate for OpenCL prevents us to assume control of the  orchestration code.
For instance, we have no control over the size of thread-groups and over
the type of memory allocated for each kernel parameter, we cannot perform optimizations that 
overlap communication with computation, among others.
Accordingly, we are not extracting the full potential of the hardware.
Nonetheless, our results are of the same order of magnitude of other GPGPU approaches that 
 supply performance results for a subset of JavaGrande's benchmark suite, namely \cite{lime}.

\section{Conclusions}	
\label{sec:conclusions}

This paper presented the SOMD programming and execution model, and how it can be realized in multiple architectures, 
namely multi-cores, GPUs, and clusters of both.
We addressed the former two in detail, elaborating  on the effective compilation process and runtime system support.

The simplicity of  expressing  data parallelism at the subroutine level 
provides a powerful framework for the parallel computing of  heterogeneous computational systems, composed of multi-core CPUs, and  accelerators, such as GPUs and Accelerator Processing Units (APUs). 
The simple annotation of Java methods with data partition and reduction strategies, and the eventual delimitation of data-centric synchronized blocks, 
is sufficient to have a method compiled for parallel execution in either architecture. 
Despite its simplicity our approach provided very good performance results,  when compared to  hand-tuned  multi-threaded  applications - 
the most common way of expressing parallelism.

In conclusion, the SOMD approach paves a possible way for the increasing adoption of data parallel heterogeneous computing in mainstream software development.

 \section*{Acknowledgement}
 This work used hardware acquired in the scope of project PTDC/EIA-EIA/102579/2008 - Problem Solving Environment for Materials Structural Characterization via Tomography funded by the  Portuguese national funding agency for science, research and technology (FCT-MEC).

\bibliographystyle{elsarticle-num}
\bibliography{main}

\end{document}